\documentclass[prd,superscriptsize,reprint
,superscriptaddress
]{revtex4-1}
\usepackage[utf8]{inputenc} 
\usepackage{graphicx,xcolor,overpic,mathtools}
\usepackage{amsthm,amsmath,amssymb}
\usepackage[colorlinks]{hyperref}
\definecolor{coolblack}{rgb}{0.0, 0.18, 0.39}
\hypersetup{
    colorlinks = false,
   }
\PassOptionsToPackage{normalem}{ulem}
\usepackage{ulem} 
\usepackage{sidenotes}
\usepackage{bm}
\usepackage{bbold}
\usepackage{url}
\usepackage{physics}
\usepackage[makeroom]{cancel}
\usepackage{cleveref}
\usepackage{tensor}
\usepackage{mathrsfs}
\usepackage{todonotes}
\usepackage{slashed}
\usepackage[mode=buildnew]{standalone}
\newcommand{\lag}{\mathcal{L}}

\usepackage{lipsum}
\usepackage{diagbox}

\NewDocumentCommand{\evat}{sO{\bigg}mm}{%
  \IfBooleanTF{#1}
   {\mleft. #3 \mright|_{#4}}
   {#3#2|_{#4}}%
}
\usepackage{enumerate}
\definecolor{azure}{rgb}{0.0, 0.5, 1.0}

\newcommand{\comment}[1]{}
\newcommand{\dmu}{\partial_\mu}
\begin{document}

\title[]{Kaon condensation in skyrmion matter and compact stars}

\author{Christoph Adam}
\email[]{adam@fpaxp1.usc.es}
\affiliation{%
Departamento de F\'isica de Part\'iculas, Universidad de Santiago de Compostela} \affiliation{Instituto
Galego de F\'isica de Altas Enerxias (IGFAE) E-15782 Santiago de Compostela, Spain
}

\author{Alberto Garc\'ia Mart\'in-Caro}
\email{alberto.martin-caro@usc.es}
\affiliation{%
Instituto
Galego de F\'isica de Altas Enerxias (IGFAE) E-15782 Santiago de Compostela, Spain
}
\affiliation{Physics Dept., Brookhaven National Laboratory, Bldg. 510A, Upton, NY 11973, USA}

\author{Miguel Huidobro}%
\email[]{miguel.huidobro.garcia@usc.es}
\author{Ricardo V\'azquez}
\email[]{vazquez@fpaxp1.usc.es}
\affiliation{%
Departamento de F\'isica de Part\'iculas, Universidad de Santiago de Compostela} \affiliation{Instituto
Galego de F\'isica de Altas Enerxias (IGFAE) E-15782 Santiago de Compostela, Spain
}%
\author{Andrzej Wereszczynski}
\email[]{andrzej.wereszczynski@uj.edu.pl}
\affiliation{
Institute of Physics, Jagiellonian University, Lojasiewicza 11, Krak\'ow, Poland
}%

\date[ Date: ]{\today}
\begin{abstract}
We address the possibility of the appearance of a charged kaon condensate in neutron star cores described within a generalized Skyrme model. Our treatment of strange degrees of freedom is based on the Bound State Approach by Callan and Klebanov, which allows to obtain an in-medium effective potential for the $s$-wave kaon condensate. We predict the onset of kaon condensation at a certain threshold density---whose value depends on the parameters of the model, and ranges between $1.5$ and $2.5$ times saturation density---, and obtain both the particle fractions and equation of state for dense matter in the kaon condensed phase. Finally, we discuss the effect of such condensates on the mass-radius curves and other observable properties of neutron stars with kaon condensed cores.
\end{abstract}
\maketitle
\tableofcontents

\section{Introduction}
While heavy ion collision experiments and lattice QCD 
simulations provide insight into the properties of hot and dense QCD, neutron stars (NS) are the only known objects in the universe that may allow us to deepen our understanding of the rich structure of cold, ultra-dense strongly interacting nuclear matter. 
The astrophysical inference of NS masses, radii and moments of inertia in low mass x-ray binaries and isolated NS from the NICER experiment as well as in binary NS inspirals and subsequent mergers from  Gravitational-Wave (GW) observatories have helped to significantly constrain
the NS equation of state (EoS) at supra-saturation densities, which cannot be reached in laboratory experiments. 

These constraints, however, do not yield any information on the microscopic structure of dense matter. Indeed, owing to the non-perturbative nature of QCD at energies below the confinement scale, the precise phase structure of cold, strongly interacting matter at both finite baryon and isospin chemical potential is still very speculative. Novel phases of dense baryonic matter are expected to occur in the inner core of NS, containing additional particle species such as $\Delta$ isobar resonances \cite{Li:2018qaw}, hyperons \cite{Glendenning:giant}, or pion or kaon condensates \cite{1975ApJ...199..471H, CELENZA197723,Kaplan:1987sc,Glendenning:1997ak,Pal:2000pb}. There have been theoretical proposals of even more exotic
scenarios, where a transition to deconfined quark matter takes place inside the core \cite{PhysRevLett.70.1355, Benvenuto:1999uk, annala2020evidence},
or a new state of matter in which both hadronic and quark degrees of freedom coexist, the so-called quarkionic matter \cite{McLerran:2007qj,McLerran:2018hbz}. Studies of the dynamical features of compact stars, such as the occurrence of phase transitions during mergers or the cooling rate of proto-neutron stars may produce complementary data, as
they may strongly depend on the specific microscopic degrees of freedom as well as on the EoS.

Many calculations for the EoS in dense matter predict that strangeness degrees of freedom may become important in the interior of compact stars, in the form of hyperons (strange baryons) or a Bose-Einstein condensate
of negatively charged kaons, for densities just a few times nuclear saturation. For a recent review, see \cite{Tolos:2020aln}. Indeed, hyperons may become stable at sufficiently high isospin chemical potential, where the decay of neutrons relieve the Fermi pressure exerted by the nucleons. On the other hand, the strong attraction between $K^-$ mesons and baryons
increases with density and lowers the energy of the zero momentum state. A condensate is formed
when this energy equals the kaon chemical potential, since kaons are favored over negatively charged fermions for achieving charge neutrality, as they are bosons and can condense in the lowest energy state.

It is generally assumed that hyperons should appear at densities above
$\sim 2 - 3$ times the nuclear saturation density $n_0$, whereas the critical density
for kaon condensation is usually predicted to be a bit larger, around $\sim(3-4)n_0$ (although the specific values are of course model and parameter dependent).
A density of this order is smaller than the central density of a typical NS, so a kaon condensate
could be present in its core. The possibility of kaon condensates in the core of neutron stars has been extensively investigated in the literature, using different approaches. Its appearance tends to soften the EoS, producing smaller values for the allowed maximum masses. Therefore, the presence of hyperons at too low densities is not compatible with the stiffness required by the existence of such massive stars. This is the so-called \emph{Hyperon puzzle}, presently a subject of very active research \cite{,Bombaci:2016xzl,Vidana:2015rsa}.
 
In this work we will make use of a Generalized Skyrme model, a phenomenological, nonlinear chiral model that, due to its nonperturbative nature, can in principle be used as a simple model to study strongly interacting matter at all scales, from single baryons and nuclei to nuclear matter in neutron stars. The model contains a rather limited number of fundamental degrees of freedom, which in the simplest version are just pions, encoded into an $SU(2)$ valued field $U$, as chiral symmetry is nonlinearly realized. Nucleons and atomic nuclei emerge as collective, topologically nontrivial excitations of the mesonic fields. Mathematically they are described by topological solitons, called skyrmions, whose topological degree can be identified with the baryon number. The most attractive feature of the model is the small number of free parameters, which implies a rather strong predictive power.

The Skyrme model \cite{skyrme1962unified} and its generalizations have been applied successfully to the description of nucleon properties \cite{adkins1983static}, \cite{Ding:2007xi}, nuclear interaction potentials \cite{Halcrow:2020gbm}, ground and excited states of atomic nuclei \cite{Lau:2014baa}, \cite{Halcrow:2016spb}, or the problem of nuclear binding energies \cite{Adam:2013wya,Gillard:2015eia,Gudnason:2016mms}. Typically, the best fit to phenomenological observations requires the extension of the original Skyrme Lagrangian \cite{skyrme1962unified}, either by the addition of new degrees of freedom, e.g., vector mesons \cite{Naya:2018kyi}, or additional, physically-motivated higher derivative terms, like the so-called sextic term \cite{Adam_2010}, an effective term related to two-body interactions mediated by $\omega$ mesons. 

Simultaneously, in the last years there has been a significant progress in the application of the Skyrme model to investigate properties of dense nuclear matter and neutron stars \cite{naya2019neutron,Adam:2020yfv,Adam:2021gbm,Adam:2022aes}, 
where the sextic term is especially important, as this part of the action provides the leading contribution in the regime of high pressure and density \cite{Adam_2015a}. Indeed, it makes the skyrmionic matter much stiffer at extreme conditions which results in physically acceptable values of the maximal mass of neutron stars.

The extension of the Skyrme model for a larger number of flavors has also been discussed in the literature. In particular, for $N_F=3$, it has been used to describe pentaquarks and strange hyperons, both in the flavor symmetric limit, in which the full $SU(3)_F$ group is quantized, and in the flavor symmetry breaking limit, the so called Bound State Approach \cite{callan1985bound,Klebanov1990}, in which oscillations into the strange sector are treated as perturbations of the $SU(2)$ valued classical soliton. Since kaon degrees of freedom are best described in the latter approach, in this paper we will take this path and extend its application to the crystalline phases of Skyrmion matter, in order to be able to describe the phenomenon of kaon condensation in dense matter as predicted by the Skyrme model. 

The paper is organized  as follows: in section II, we introduce the model and review the classical and quantum properties of crystal solutions. In section III we review the bound state approach to kaons in the Skyrme model, and compute the contribution to the total energy from the kaon condensate. In section IV, we obtain the system of equations from minimizing the free energy of $npe\mu\bar{K}$ matter, and solve it to find the onset of kaon condensation for different sets of parameters, and finally in section V we calculate the Equation of State of skyrmion matter including a kaon condensate, and compute the corresponding  mass-radius curves for NS.

\section{Generalized Skyrme model and skyrmion crystals}

The generalized Skyrme model we will consider is given by the following Lagrangian density
\begin{align}
    \notag \lag = -\frac{f^2_{\pi}}{16}&\Tr L_{\mu}L^{\mu}  + \frac{1}{32e^2}\Tr \left[L_{\mu},L_{\nu}\right]^2  \\[2mm] 
    &- \lambda^2 \pi^4\mathcal{B}_{\mu}\mathcal{B}^{\mu} + \frac{m^2_{\pi} f^2_{\pi}}{8}\Tr \left(U - I \right),
    \label{Lag}
\end{align}
where $L_\mu=U^\dagger \partial_\mu U$ is the left invariant Maurer-Cartan current and the Skyrme field can be written as
\begin{equation}
U = \sigma + i \pi_a \tau_a.
 \label{Ufield}
\end{equation}
Here, $\pi_a$ ($a$ = 1, 2, 3) are the pions and $\tau_a$ are the Pauli matrices. The unitarity of the matrix field implies $\sigma^2 + \pi_a\pi_a = 1$. 
Furthermore, $\mathcal{B}^\mu$ is the conserved topological current which, in the standard manner, defines the topological index of maps $U$, i.e., the baryon charge $B$
\begin{equation}
    B = \int d^3x \mathcal{B}^0, \hspace{2mm} \mathcal{B}^{\mu} =\frac{1}{24\pi^2} \epsilon^{\mu\nu\alpha\beta}\Tr\left\{ L_{\nu}L_{\alpha}L_{\beta} \right\}.
    \label{TopoNumber}
\end{equation}
The generalized Skyrme effective model contains only four terms and, therefore, four coupling constants, $f_\pi, m_\pi, e, \lambda$,  two of which have a direct phenomenological interpretation as the pion decay constant and the pion mass. In addition, $\lambda$ can be related to a ratio between the mass and the coupling constant of the $\omega$ meson. From the very beginning, we assume the physical mass of the pions, $m_\pi=140$ MeV. The remaining constants are fitted to some properties of infinite nuclear matter.

We remark that the third term in the action, although often omitted in the context of light nuclei, is obligatory when one studies the properties of nuclear matter at high density, which is a natural environment in the core of neutron stars. Indeed, this sextic term governs the equation of state at this regime and asymptotically leads to the maximally stiff EoS. 


The canonical description of NS is provided by the relativistic TOV approach where a particular model of infinite nuclear matter gives a source term of the Einstein equations. Effectively, it enters via an EoS, that is, a relation between e.g., pressure and density. In the Skyrme model, infinite skyrmionic matter is described by a periodic minimizer of the static energy ($E = -\int d^3x \lag$) and, therefore, it is usually referred to as the Skyrme crystal. Obviously, while the total energy of the crystal is infinite, the energy per baryon number remains finite
\begin{equation}
    \frac{E}{B} = \frac{N_{\text{cells}}\:E_{\text{cell}}}{N_{\text{cells}}\:B_{\text{cell}}}=\frac{E_{\text{cell}}}{B_{\text{cell}}}.
\end{equation}
Here, $N_\text{cells}$ is the number of cells and $E_\text{cell}$, $B_\text{cell}$ are the energy and baryon charge in a single, periodic cell. 
As mentioned before, skyrmion crystals minimize the static energy functional,
\begin{eqnarray}
     E &=& 
    \int d^3 x \left(\mathcal{E}_2 + \mathcal{E}_4 + \mathcal{E}_6 + \mathcal{E}_0 \right) \nonumber \\  
    &=& \frac{1}{24\pi^2} \int d^3x \left[ -\frac{1}{2}\Tr \{L_iL_i\} - \frac{1}{4}\Tr \{\left[L_i,L_j\right]^2\}  \right. \nonumber \\
    && + \left.  4\pi^4 c_6 (\mathcal{B}^0)^2 + \frac{c_0}{2}\Tr\left( I - U \right)  \right] \label{Energy} 
\end{eqnarray}
over a finite region of space with periodic boundary conditions. Here $U$ is the SU(2) valued Skyrme field and $L_\mu=U^\dagger \partial_\mu U$. Further, $\mathcal{E}_2$ and $\mathcal{E}_4$ are the standard terms of the Skyrme model quadratic and quartic in derivatives, and $\mathcal{E}_6$ is the sextic term mentioned above.
Finally, $\mathcal{E}_0$ is the pion mass potential. We have defined the dimensionless constants $c_6 = 2\lambda^2 \frac{f^2_{\pi} e^4}{\hbar^3}$, ${c_0 = 2\frac{m^2_{\pi}}{f^2_{\pi}e^2}}$ and use the so-called Skyrme model units, so that our energy and length units are
\begin{equation}
    E_s = 3\pi^2 f_{\pi}/e,\quad l_s = \hbar/(f_{\pi}e),
\end{equation}
respectively. Both the size of the unit cell (characterized by the unit cell length parameter $L$) and its geometry will affect $E_{\rm cell}$.
It turns out that, for our purposes, the ground state of skyrmion crystals is well described by a cubic unit cell with side length $2L$ composed of skyrmions in a face-centered cubic (FCC) arrangement, but with an additional symmetry. Concretely, it respects the following symmetries,
\begin{align}
    \notag\text{S}_1&: (x,y,z) \rightarrow (-x,y,z), \\[2mm] &(\sigma,\pi_1,\pi_2,\pi_3) \rightarrow (\sigma,-\pi_1,\pi_2,\pi_3), \\[2mm]
    \notag\text{S}_2&: (x,y,z) \rightarrow (y,z,x), \\[2mm] &(\sigma,\pi_1,\pi_2,\pi_3) \rightarrow (\sigma,\pi_2,\pi_3,\pi_1),\\[2mm]
    \notag\text{S}_3&: (x,y,z) \rightarrow (x,z,-y), \\[2mm] &(\sigma,\pi_1,\pi_2,\pi_3) \rightarrow (\sigma,-\pi_1,\pi_3,-\pi_2), \\[2mm]
    \notag\text{S}_4&: (x,y,z) \rightarrow (x+L,y,z), \\[2mm] &(\sigma,\pi_1,\pi_2,\pi_3) \rightarrow (-\sigma,-\pi_1,\pi_2,\pi_3).
\end{align}
A detailed description of the construction of the Skyrme crystal and the comparison of different symmetries can be found in \cite{Adam:2021gbm}. As in that previous work, the unit cell that we will consider has size $2L$ and a baryon content of $B_{\text{cell}} = 4$. Because of the additional symmetry for this crystal, the unit cell of size $2L$ decomposes into 8 cubes of side length $L$, each forming a simple cubic arrangement of half-skyrmions, where half-skyrmions are located in the corners of the cube and lead to a baryon content of $1/2$. The fields, however, are periodic only in $2L$, hence the unit cell has side length $2L$. 
We obtain the value of the energy for each value of $L$ and, as explained in \cite{Adam:2021gbm}, the energy-size curve, $E_{\text{cell}}(L)$, is a convex function which has a minimum at a certain $L_0$.
We identify this point with the nuclear saturation point of infinite, symmetric nuclear matter, which also presents a minimum in the energy per baryon number curve as a function of the baryon density $n_B=(4/(2L)^3)= (1/2)L^{-3}$.

Up to now, we have considered the classical skyrmion crystal, which corresponds to symmetric nuclear matter.
However, for a realistic description of nuclear matter inside a NS we need to consider an almost completely isospin-asymmetric state, where only a small amount of protons is allowed.
We have already considered this scenario in \cite{Adam:2022aes} via a semiclassical quantization of the isospin degrees of freedom (DOF) of the skyrmion crystal.
Indeed, it is standard in nuclear physics to define the binding energy of a nuclear system in the following way,
\begin{equation}
    \frac{E}{B}(n_B,\delta)=E_N(n_B)+S_N(n_B)\delta^2+\order{\delta^3},
\end{equation}
where $\delta$ is the isospin asymmetry parameter, defined in terms of the proton fraction $\gamma$ of the system, $\delta=(1-2\gamma)$. $E_N(n_B)$ denotes the binding energy of isospin-symmetric matter, and $S_N(n_B)$ represents the so-called \emph{symmetry energy}, which is responsible for the change in the binding energy when the neutron-to-proton ratio changes for a fixed value of the baryon number. The knowledge of the symmetry energy at high densities is fundamental for a correct description of NS interiors. However, although the values of the symmetry energy at saturation is well known ($S_0\sim 30$ MeV) \cite{FiorellaFantina:2018dga}, the difficulty in experimentally measuring its behavior at high densities forces us to express it as an expansion in powers of the baryon density around $n_0$,
\begin{equation}
    S_N(n_B)=S_0 + \frac{n_B-n_0}{3n_0}L_{\rm sym} +\frac{(n_B-n_0)^2}{18n_0^2} K_{\rm sym} +\cdots
\end{equation}
where
\begin{equation}
    L_{\rm sym}=3 n_0\pdv{S_N}{n_B}\eval_{n_B=n_0}, \quad  K_{\rm sym}=9n_0^2\pdv[2]{S_N}{n_B}\eval_{n_B=n_0}
    \label{eq_symetobs}
\end{equation}
denote the slope and curvature of the symmetry energy at saturation, respectively. The values of these coefficients are still very uncertain, but recent analysis of combined astrophysical and nuclear observations made possible to constrain the symmetry energy above $n_0$ \cite{Landry:2021ezp,Tang:2021snt,deTovar:2021sjo,Gil:2021ols,Li:2021thg}.

Let us now review the procedure for calculating the symmetry energy of an $SU(2)$ skyrmion crystal, as it will be generalized to the 3 flavor case in the next section.
\label{sec:Isospin}
First, let us rewrite the Skyrme Lagrangian \eqref{Lag} as
\begin{equation}
    \mathcal{L}=a\Tr\left\{L_{\mu}L^{\mu}\right\} + b\Tr\left\{\left[L_{\mu},L_{\nu}\right]^2\right\} + c\,\mathcal{B}_{\mu}\mathcal{B}^{\mu} + d\Tr \left(U - I \right).
\end{equation}
In our dimensionless units, we have 
\begin{equation}
    a=-\frac{1}{2}, \quad b=\frac{1}{4},\quad c=-8\lambda^2\pi^4f_\pi^2e^4, \quad 
    d = \frac{m^2_{\pi}}{f^2_{\pi}e^2}.
\label{adimpars}
\end{equation}
and consider a (time-dependent) isospin transformation of a static Skyrme field configuration:
\begin{equation}
    U(\vec{x})\rightarrow \tilde U(\vec{x},t)\equiv g(t) U(\vec{x}) g^\dagger(t).
    \label{IsorotF}
\end{equation}
The time dependent isospin matrices $g(t)$ are collective coordinates, whose dynamics  is given by a kinetic term in the energy functional,
\begin{equation}
    T=\frac{1}{2}\omega_i\Lambda_{ij}\omega_j
\end{equation}
where $\Lambda_{ij}$ is the isospin inertia tensor, given by
\begin{eqnarray}
    \Lambda_{ij}& =& \int \{2a\Tr\{T_iT_j\}-4b\Tr\{[T_i,L_k][T_j,L_k]\}- \nonumber \\
    &&-\frac{c}{32\pi^4}\varepsilon^{abc}\Tr\{T_iL_bL_c\}\varepsilon_{ars}\Tr\{T_jL_rL_s\}\}\,d^3x \nonumber \\
    &=& \; \Lambda \, \delta_{ij}
    \label{Inertia_Tensor}
\end{eqnarray}
being $ T_a$ the $\mathfrak{su}(2)$-valued current
$
    T_a=\frac{i}{2}U^\dagger[\tau_a,U]
$
and $\vec{\omega}$ the associated isospin angular velocity, defined by $g^\dagger\dot{g}=\tfrac{i}{2}\omega_a\tau_a$.

As shown in \cite{Adam:2022aes}, we may canonically quantize the isospin collective degrees of freedom and obtain a Hamiltonian, which for a cubic crystal with a number $N$ of unit cells is given by 
\begin{equation}
    H=\frac{\hbar^2}{2N\Lambda_{\rm cell}}I^{\rm{tot}}(I^{\rm{tot}}+1)
\end{equation}
in terms of the isospin moment of inertia $\Lambda = N\Lambda_{\rm cell}$, and the total isospin angular momentum eigenvalue $I^{\rm tot}$, given by the product of the total number of unit cells times the total isospin of each unit cell, which can be obtained by composing the isospins of each of the cells. In the charge neutral case, all cells will have the highest possible value of isospin angular momentum, so that on each unit cell with baryon number $B$, the total isospin will be $\frac{1}{2}B$, and hence for the full crystal will be $I^{\rm{tot}}=\frac{1}{2}N B$.

Thus, the quantum correction to the energy (per unit cell) due to the isospin degrees of freedom in the neutral (i.e. purely neutronic) limit would be (assuming $N\rightarrow \infty$):
\begin{equation}
    E^{\rm{iso}}=\frac{\hbar^2}{8\Lambda_{\rm cell}}B^2,
    \label{Eiso}
\end{equation}
where the value of $\hbar$ is related to the value of $e$ through:
\begin{equation}
    \hbar = \frac{e^2}{3\pi^2}.
\end{equation}
The classical skyrmion crystal configurations can therefore be understood as models for isospin-symmetric nuclear matter, i.e. nuclear matter with zero total isospin.
Deviations from the exact isospin symmetric case yield quantum isospin corrections to the crystal energy per baryon, which depend on the difference between protons and neutrons through the total isospin number per unit cell. Hence, by considering the effect of iso-rotations over classical solutions we are effectively breaking the isospin symmetry of the static energy functional by adding a correction of quantum origin that explicitly breaks it. 
Moreover, knowing the energy correction due to isospin, it is straightforward to obtain the associated isospin chemical potential for the skyrmion crystal using its thermodynamical definition:
$
    \mu_I=-\pdv{E}{N_I}
$, 
where $N_I$ is the (third component of) the isospin number per unit cell. Given that $(I^{\rm{tot}})^2=I_1^2+I_2^2+I_3^2$ and $N_I=I_3/N$, we may rewrite \cref{Eiso} as
\begin{equation}
    E^{\rm{iso}}=\frac{1 }{2\Lambda_{\rm cell}}\qty(N_I^2+\frac{I_2^2}{N^2}+\frac{I_1^2}{N^2})
\end{equation}
and then
\begin{equation}
    \mu_I=-\pdv{E^{\rm{iso}}}{N_I}=-\frac{N_I}{\Lambda_{\rm cell}}.
\end{equation}

Let us now consider a finite chunk of the Skyrme crystal of $N$ unit cells, and let $A=N\times B_{\rm cell}$, where $B_{\rm cell}$ is the baryon number of a unit cell. We do not enforce charge neutrality at this step, and further leave unknown the quantum state of the crystal. 
As in \cite{Adam:2022aes}, we take  a mean field approximation and consider that the isospin density in an arbitrary skyrmion crystal quantum state is approximately uniform so that
\begin{equation}
    \ev{I^0_3}=\frac{\ev{I_3}}{\int d^3x}=\frac{\ev{I_3}}{NV_{\rm cell }}\doteq \frac{N_I}{V_{\rm cell}}
\end{equation}
where $N_I$ is the isospin charge  per unit cell in this arbitrary quantum state. The effective proton fraction that would yield such an isospin charge per unit cell is $\gamma$, so we write
    \begin{equation}
       N_I=-\frac{1}{2}(1-2\gamma)B_{\rm cell}=-\frac{B_{\rm cell}}{2}\delta .
    \end{equation}
Hence, the isospin energy per unit cell of the skyrmion crystal in such a state can be written in terms of the asymmetry parameter
\begin{equation}
    E^{\rm iso}=\frac{\hbar^2}{8\Lambda}B_{\rm cell}^2\delta^2,
    \label{isospin_energy}
\end{equation}
and thus the symmetry energy for Skyrme crystals is given by
\begin{equation}
    S_N(n_B)=\frac{L^3}{8\Lambda}n_B.
\end{equation}
As argued in \cite{klebanov1985nuclear}, any quantum state different from purely neutron matter leads to a divergent Coulomb energy term for the skyrmion crystal. Therefore, in order to allow for a nonzero positive electric charge within the unit cell we consider the existence of a neutralizing background of negatively charged leptons, namely, electrons and muons. In this scenario, the effects of the positive charge become almost completely screened, and the residual Coulomb energy is negligible, so we do not take it into account.

Apart from electromagnetic forces, nuclear matter interacts with leptons via the weak force. Indeed, the exchange between leptons and nucleons inside NS is completely described by the $\beta$-decay and electron capture processes,
\begin{equation}
    n\rightarrow p+l+\bar{\nu}_l\quad ,\quad p+l\rightarrow n+\nu_l,
    \label{eq:URCA}
\end{equation}
which take place simultaneously, as long as the \textit{charge neutrality} and $\beta$\textit{-equilibrium} conditions are satisfied,
\begin{align}
    n_p&=\frac{Z}{V}=n_e+n_\mu, \\[2mm]
    \mu_n=\mu_p+\mu_l &\implies\mu_I=\mu_l,\quad l=e,\mu.
    \label{betacond}
\end{align}
Leptons inside a NS can be described as a non-interacting relativistic Fermi gas. Then the chemical potential for a given kind of lepton is,
\begin{equation}
    \mu_l=\sqrt{(\hbar k_{F})^2+m_l^2},
\end{equation}
where $k_{F}=(3\pi^2n_l)^{1/3}$ is the corresponding Fermi momentum, and $m_l$ is the mass of the corresponding lepton. Considering the most general case, in which we include muons, we combine this last expression with the above equilibrium conditions and obtain the following system of equations,
\begin{align}
    n_\mu = \frac{1}{3\pi^2}&\qty[\qty(\frac{\hbar B_{\rm cell}(1-2\gamma)}{2\Lambda})^2-\left(\frac{m_\mu}{\hbar}\right)^2]^{\tfrac{3}{2}}, \\[2mm]
    \frac{\hbar B_{\rm cell}}{2\Lambda}&(1-2\gamma) = \qty[3\pi^2\qty(\frac{\gamma B_{\rm cell}}{8L^3}-n_\mu)]^{\tfrac{1}{3}},
\end{align}
In order to solve the system for $\gamma$, we take the ultrarelativistic approximation $\mu_e \approx \hbar k_{F, e}$ for electrons. Besides we start solving the system at low densities considering only electrons, hence we drop the first equation and set $n_{\mu} = 0$ until the condition $\mu_e = m_{\mu}$ is reached. Then, muons start to appear and we solve both equations. For each length of the unit cell we obtain the value of the proton fraction, hence we reconstruct the curve $\gamma(L)$.

The total energy per unit cell in a $\beta$-equilibrated skyrmion crystal is therefore given by
\begin{equation}
    E= E_{\rm class} + E_{\rm iso}(\gamma)+E_{e}(\gamma)+E_{\mu}(\gamma),
    \label{npemu_matter}
\end{equation}
where $E_{\rm class}$ correspond to the classical energy of the Skyrme crystal, $E_{\rm iso}$ is calculated from \cref{isospin_energy} and the energies of the leptons are the usual energy of a relativistic Fermi gas with mass $m_l$ at zero temperature,
\begin{align}
    E_{\rm lep}&=\int_0^{k_f}\frac{k^2dk}{\pi^2}\sqrt{k^2+m_{l}^2}=\\
    &=\frac{m^4_l}{8\pi^2}\qty[x_r(1+2x_r^2)\sqrt{1+x_r^2}-\ln{x_r+\sqrt{1+x_r^2}}],\notag
\end{align}
where $x_r=k_F/m_l$. Recall that for electrons we take the approximation $x^{-1}_r \rightarrow 0$, which is justified for densities $n \geq n_0$.

\section{Kaon condensate in skyrmion crystals}
Having obtained the proton fraction (hence the electron chemical potential) in $npe\mu$-matter as a function of density, we can now turn to the question of whether kaon fields may condense inside a Skyrme crystal for a sufficiently high density, and, if so, whether this critical density value is relevant for the description of matter inside compact stars.
\subsection{Kaon fluctuations in the Skyrme model}
Following the bound-state approach first proposed in \cite{callan1985bound} we may include strange degrees of freedom in the Skyrme model by extending the skyrmion field to a $SU(3)$-valued field $U$ through modelling kaon fluctuations on top of a $SU(2)$ skyrmion-like background $u$. With the only requirement that unitarity must be preserved, different ansätze have been proposed in the literature for the total $SU(3)$ field describing both pions and kaons. In this work, we choose the ansatz proposed by Blom et al in \cite{Blom1989HyperonsAB}:
\begin{equation}
    U = \sqrt{U_K}U_{\pi}\sqrt{U_K}.
    \label{BDR ansatz}
\end{equation}
In this ansatz $U_{\pi}$ represents the $SU(3)$ embedding of the purely pionic part $u$, and the field $U_K$ are the fluctuations in the strange directions. It can be shown that this ansatz is equivalent to the one first proposed by Callan and Klebanov in \cite{callan1985bound} when computing static properties of hyperons, although both may differ in other predictions of the model \cite{nyman1990low}.

In the simplest $SU(3)$ embedding, the $SU(2)$ field $u$ is extended to $U_\pi$ by filling the rest of entries with ones in the diagonal and zeros outside. On the other hand, the kaon ansatz is modelled by a $\mathfrak{su}$(3) -valued matrix $\mathcal{D}$ which is non trivial in the off-diagonal elements:
\begin{equation}
\begin{split}
    U_{\pi} &= \begin{pmatrix}
                u & 0\\
                0 & 1
              \end{pmatrix}, \hspace{3mm}
    U_K = e^{ i \frac{2\sqrt{2}}{f_{\pi}}
                   \mathcal{D} },\\
&u=\sigma+i\pi_a\tau_a,\quad \mathcal{D}=\mqty( 0 & K\\ K^{\dagger} & 0)
\label{Kparam}
\end{split}
\end{equation}
where $K$ consists of a scalar doublet of complex fields representing  charged and neutral kaons:
\begin{equation}
    K=\mqty(K^+\\K^0),\quad K^\dagger = (K^-, \bar{K}^0).
\end{equation}
The extension of the Generalized Skyrme Lagrangian from \eqref{Lag} to include strange degrees of freedom consists in the replacement of the $\lag_0$ term by \cite{nyman1990low}:
\begin{align}
    \notag \lag_0^{\rm new} =&  \frac{f^2_{\pi}}{48}\left( m^2_{\pi} + 2 m^2_K \right) \Tr{U + U^{\dagger} - 2} +\\ &+\frac{\sqrt{3}}{24}f^2_{\pi}\left( m^2_{\pi} - m^2_K \right)\Tr{\lambda_8\left(U + U^{\dagger}\right)},
\end{align}
where $\lambda_8$ is the eighth Gell-Mann matrix and $m_K$ is the vacuum kaon mass, and the addition of the Wess-Zumino-Witten (WZW) term, which can be expressed in terms of a 5-dimensional action:
\begin{equation}
    S_{WZ} = -i\frac{N_c}{240\pi^2}\int d^5x \: \epsilon^{\mu\nu\alpha\beta\gamma}\Tr{L_{\mu}L_{\nu}L_{\alpha}L_{\beta}L_{\gamma}}.
\end{equation}

\subsection{The kaon condensate on classical crystals }
The onset of kaon condensation in the Skyrme model takes place at a critical density $n_{\rm cond}$ at which $\mu_{e}$ becomes greater than the energy of the kaon zero-momentum mode (s-wave condensate). Thus, for baryon densities $n\geq n_{\rm cond}$, the macroscopic contribution of the kaon condensate to the energy must be taken into account when obtaining the EoS of dense matter. To do so, we follow the standard procedure to describe Bose-Einstein condensation of a (complex) scalar field (see eg \cite{Schmitt:2010pn}) in which the field condensates correspond to the non-zero vacuum expectation values (vev), $\ev{K^\pm}$, which are assumed to be constant in space and whose time dependence is given by:
\begin{equation}
    \ev{K^\mp}=\phi e^{\mp i\mu_K t}
\end{equation}
The real constant $\phi$ corresponds to the zero-momentum component of the fields, which acquires a nonvanishing, macroscopic value after the condensation. Its exact value is determined from the minimization of the corresponding effective potential, to whose calculation we will dedicate the rest of this section. On the other hand, the phase $\mu_K$ is nothing but the corresponding kaon chemical potential. First, we will need an explicit form of the $SU(3)$ Skyrme field in the kaon condensed phase. Assuming the charged kaons will be the first mesons to condense \footnote{Actually, that the charged (in particular, the negatively charged) kaons will condense first is true in our approach (whenever $\mu_e>0$), since the chemical potential associated to neutral kaons is zero, so that the onset of neutral kaon condensation is given by $m^*_K=0$.}, we can safely drop the neutral kaon contribution, and define the following matrix
\begin{equation}
    \tilde{\mathcal{D}}=\mqty(0&0&\phi e^{i\mu_K t}\\
    0&0&0\\
    \phi e^{-i\mu_k t}&0&0)
\end{equation}
which results from substituting the kaon fields in $\mathcal{D}$ as defined in \eqref{Kparam} by their corresponding vev in the kaon condensed phase. Also, taking advantage of the property $\mathcal{D}^3=\phi^2\mathcal{D}$, we may write the $SU(3)$ element generated by $\tilde{\mathcal{D}}$ explicitly in matrix form:
\begin{equation}
    \Sigma=e^{i \tfrac{\sqrt{2}}{f_\pi}\tilde{\mathcal{D}}}=\mqty(\cos\tilde{\phi}&0&ie^{i\mu_K t}\sin \tilde\phi\\
    0&1&0\\
    ie^{-i\mu_K t}\sin \tilde\phi&0&\cos\tilde\phi)
    \label{Sigmaexpl}
\end{equation}
where $\tilde\phi=\tfrac{\sqrt{2}}{f_\pi}\phi$ is the dimensionless condensate amplitude.

Furthermore, assuming the backreaction from the kaon condensate to the skyrmion crystal is negligible, and thus the classically obtained crystal configuration will be the physically correct background even in the kaon condensed phase, we may write the $SU(3)$ field in this phase as $U=\Sigma U_\pi \Sigma$, where $U_\pi $ is the $SU(3)$ embedding of the $SU(2)$ skyrmion background as in \eqref{Kparam}. Introducing this $U$ in the total action yields the standard Skyrme action for the $SU(2)$ field plus an effective potential term for the kaon condensate:
\begin{equation}
    S_{Sk}(U)+S_{\rm WZW}(U)=S_{Sk}(U_\pi)-\int d t V_K(\tilde{\phi}),
\end{equation}
where
\begin{equation}
    V_K = \frac{1}{24\pi^2}\int d^3x\Big[ V^{(2)}_K + V^{(4)}_K + V^{(6)}_K + V^{(0)}_K\Big] + V^{(WZW)}_K .
\end{equation}
Let us now calculate the contribution to the effective potential $V_K$ of each term in the action:
\begin{itemize}
    \item Quadratic term:
    Given that the crystal background is static and the kaon condensate does not depend on spatial coordinates, the kaon part of the quadratic term may be written as
    \begin{equation}
        \Tr{L_0^2}= -[\Tr\{\partial_t \Sigma^\dagger \partial_t \Sigma\}+\Tr\{ \Sigma^\dagger\partial_t \Sigma U_\pi^\dagger \Sigma \partial_ t\Sigma^\dagger U_\pi\}].
    \end{equation}
    Introducing the explicit expression for $\Sigma$, \eqref{Sigmaexpl}, yields
    \begin{equation}
        V_K^{(2)}=\mu_K^2\sin^2\tilde{\phi}[(1+\sigma^2+\pi^2_3)\sin^2\tilde\phi-2(1+\sigma\cos^2\tilde\phi)].
    \end{equation}
    
    \item Quartic term:
    In the quartic term, the kaon effective potential comes from the terms with time derivatives of the total field,
    \begin{align}
        &\Tr{[L_0,L_i]^2}=2[\Tr\{\partial_t U^{\dagger}\partial_i U \partial_t U^\dagger\partial_i U\}-\\[2mm]
        &-\Tr\{\partial_i U^\dagger\partial_t U \partial_i U^\dagger\partial_t U\}],\notag
    \end{align}
    which, after substitution of the expression for $\Sigma$, gives
    \begin{align}
      V_K^{(4)}&= -2\mu_K^{2}\sin^{2}{\tilde\phi} \big\{(1+\sigma)\partial_i n^2\cos^{2}{\tilde\phi} + \\[2mm]
      + & 2[ \partial_ i \sigma^{2} (1-\pi_{3}^{2}) +  \partial_i \pi_3^{2} (1-\sigma^{2}) + 2\sigma \pi_{3} \partial_ i \sigma \partial_i \pi_3]\sin^2\tilde{\phi} \big\} \nonumber
    \end{align}
    
    \item Mass term:
    The kaon part associated to the mass term gives the following contribution,
    \begin{equation}
       V_K^{(0)}(\tilde\phi)= 2\frac{m^2_K}{f^2_{\pi}e^2}(1+\sigma)\sin^2\tilde\phi
    \end{equation}
    
    \item Wess-Zumino-Witten term:
    The WZW term is written as a 5-form integrated over an auxiliar 5-dimensional disk $D$ whose boundary is the spacetime manifold $M$, but in \cref{WZW-contribution} we show that the variation after the kaon fluctuations of the pion background yields a local term which may be written as an effective four-dimensional lagrangian. Indeed, we show that 
    \begin{align}
        &S_{\rm WZW}(U)=\quad S_{\rm WZW}(U_\pi)+\notag\\
        &-\frac{i N_C}{2}\int_M B^0\Tr{\mqty(\mathbb{1}_2&0\\0&0)\qty(\Sigma\partial_t \Sigma^\dagger+U_\pi^\dagger \Sigma ^\dagger\partial_t\Sigma U_\pi )}=\notag\\
        &=-N_C\int \mu B_{\rm cell} \sin^{2}{\left(\phi \right)}dt=\int V_K^{\rm (WZW)}(\phi) dt.\label{WZWkaoncontrib}
    \end{align}
    
    \item Sextic term:
    The contribution from the sextic term is also obtained in \cref{WZW-contribution} to be
\begin{equation}
  V_K^{(6)}=-\frac{\lambda^2 f^2_{\pi} e^4}{16} \Tr\{ [R_j, R_k]\xi_0 \}^2
\end{equation}
where $\xi_\mu=U_\pi \Sigma\partial_\mu\Sigma^\dagger U_\pi^\dagger-\Sigma^\dagger \partial_\mu \Sigma$. Once the traces are evaluated, we end up with
\begin{align}
    V_K^{(6)}= &- \lambda^2 f^2_{\pi} e^4 \mu_K^2 \sin^4(\tilde\phi) (\partial_i\pi_3\partial_j\sigma-\partial_i\sigma\partial_j\pi_3)^2.
\end{align}

\comment{
\begin{align}
    V_K^{(6)}= &-4 \lambda^2 f^2_{\pi} e^4 \mu_K^2 \sin^4(\tilde\phi) \left[ -\vec{\pi} \partial_i \vec{\pi} \times \partial_j \vec{\pi} \pi_3 +\right. \\
    &\left.+2\partial_{[i} \pi_1 \partial_{j]} \pi_2 (1-\sigma^{2}) + \partial_{[i} \sigma \partial_{j]} \pi_3 (1-2\sigma^{2})
         \right]^2.\notag
\end{align}
}
\end{itemize}

\vspace*{0.5cm}
\subsection{ Effect of kaon condensation on the quantum corrections to Skyrme crystals}
In the above calculations, we have taken separately the contributions of a kaon condensate and an isospin angular momentum of the skyrmion crystal, and the kaon condensate interacts with the skyrmion isospin only indirectly via the charge neutrality and $\beta$ equilibrium conditions, which relate their corresponding chemical potentials. However, since kaons possess an isospin quantum number, we should consider a (time-dependent) isospin transformation of the full Skyrme field + kaon condensate configuration $U=\Sigma U_\pi \Sigma$:
\begin{equation}
    U\rightarrow \tilde U\equiv A(t) U A^\dagger(t),
    \label{IsorotF}
\end{equation}
where $A$ is an element of $SU(3)$ modelling an isospin rotation,
\begin{equation}
    A=\mqty(a&0\\0&1), \quad a\in SU(2).
\end{equation}
The Maurer-Cartan form transforms as ($\dot{A}=dA/dt$)
\begin{equation}
    \Tilde{U}^\dagger\partial_\mu \Tilde{U}=\left\{\begin{array}{ll}
         A U^\dagger\partial_i U A^\dagger,\quad (\mu=i=1,2,3), &  \\[2mm]
        A U^\dagger\partial_0 U A^\dagger+ A(U^\dagger[A^\dagger \dot{A},U])A^\dagger,&(\mu=0). 
    \end{array}\right.
    \label{MCiso}
\end{equation}
We now define the isospin angular velocity $\vec{\omega}$ as $A^\dagger\dot{A}=\tfrac{i}{2}\omega_a\lambda_a$ ($a=1,2,3$),
with $\lambda_\textsc{a}$ the Gell-Mann matrices generating $SU(3)$ for $\textsc{a}=1,\cdots 8$. Notice that $\vec{\omega}$ is a three-vector, since $A^\dagger\dot{A}$ belongs to the isospin $\mathfrak{su}(2)$ subalgebra of $\mathfrak{su}(3)$
Then, we may write the time component of the Maurer-Cartan current as
$ \tilde U^\dagger\partial_0 \tilde U=AL_0A^\dagger +A T_a A^\dagger \omega_a$, where $ T_a$ is the $\mathfrak{su}(3)$-valued current:
\begin{equation}
    T_a=\frac{i}{2}U^\dagger[\lambda_a,U]\equiv iT_a^\textsc{a}\lambda_\textsc{a},
    \label{UdcommtU}
\end{equation}
where we have made use of the parametrization \eqref{Kparam}. 
\comment{
Moreover, the spatial components of the M-C form can be written in terms of the sigma and pion fields as well:
\begin{equation}
    L_k=(\sigma-i\pi_a\lambda_a)(\partial_k\sigma+i\partial_k\pi_b\lambda_b)=i(\sigma\partial_k\pi_c-\pi_c\partial_k\sigma+\pi_a\partial_k\pi_b\varepsilon_{abc})\lambda_c\equiv L_k^c\lambda_c.
    \label{LeftMC}
\end{equation}
}
The time dependence of the new Skyrme field induces the existence of a kinetic term in the energy functional, given by \footnote{Remember that we are using the mostly minus convention for the metric signature.}

\begin{equation}
\begin{split}
    T=\int\{a&\qty(\Tr\{L_0L_0\}+2\Tr\{L_0T_a\}\omega_ a+\Tr\{T_aT_b\}\omega_a\omega_b)\\
    -2b\big(\Tr\{[&(L_0+T_a\omega_a),L_k][(L_0+T_b\omega_b),L_k]\}\big)-c\,\,\mathcal{B}^i\mathcal{B}_i\}d^3x,
\end{split}
\end{equation}
with $\mathcal{B}^i$ the spatial components of the topological current \eqref{TopoNumber}:
\begin{equation}
    \mathcal{B}^i=
    \frac{3}{24\pi^2}\varepsilon^{ijk}\Tr\{(L_0+T_a\omega_a)L_jL_k\}.
\end{equation}
We may rewrite the kinetic isorotational energy in the standard way as a quadratic form acting on the components of the isospin angular velocity,
\begin{equation}
    T=\frac{1}{2}\omega_a\Lambda_{ab}\omega_b+\Delta_a\omega_a - V_K
\label{kinetic}
\end{equation}
where $\Lambda_{ab}$ is the isospin inertia tensor and $\Delta_a$ is the kaon condensate isospin current, given by

\begin{widetext}
\begin{align}
    \Lambda_{ab}&=\int\left\{2a\Tr\{T_aT_b\}-4b\Tr\{[T_a,L_k][T_b,L_k]\}-\frac{c}{32\pi^4}\varepsilon^{lmn}\Tr\{T_aL_mL_n\}\varepsilon_{lrs}\Tr\{T_jL_rL_s\}\right\}\,d^3x,
    \label{Inertia_Tensor}\\
    \Delta_a&=\int\left\{2a\Tr\{L_0T_a\}-4b\Tr\{[T_a,L_k][L_0,L_k]\}-\frac{c}{32\pi^4}\varepsilon^{lmn}\Tr\{L_0L_mL_n\}\varepsilon_{lrs}\Tr\{T_aL_rL_s\}\right\}\,d^3x,
\end{align}
\end{widetext}
where $a,b$ and $c$ are those in \cref{adimpars}.

The symmetries of the crystalline configuration that we consider in this work, concretely the $S_1$ and $S_2$ transformations, imply that the isospin inertia tensor becomes proportional to the identity, i.e. $\Lambda^{\rm{crystal}}_{ab}=\Lambda\delta_{ab}$. However, the presence of a kaon condensate breaks this symmetry to a $U(1)$ subgroup, so that $\Lambda_{ab}$ presents two different eigenvalues in the condensate phase, $\Lambda_{\rm cond}=\text{diag}(\Lambda,\Lambda,\Lambda_3)$. Similarly, $\Delta_a=0$ in the purely barionic phase, and its third component acquires a non-zero value in the condensate phase, $\Delta_{\rm cond}=(0,0,\Delta)$. The explicit expressions for $\Lambda_3$ and $\Delta$ in the condensed phase are written in \cref{app2}. One can easily check that in the non-condensed phase, $\phi=0$ and the results of the previous section are recovered, namely, $\Lambda_3=\Lambda$, $\Delta=0$.

The quantization procedure now goes along the same lines as in the first section. However, the isospin breaking due to the kaon condensate implies that the canonical momentum associated to the third component of the isospin angular velocity will now be different, and given by
$ I_3=\Lambda_3\omega_3+\Delta$.

Thus, after a Legendre transformation to rewrite \eqref{kinetic} in Hamiltonian form, and making the $N\rightarrow\infty $ approximation, one can write the quantum energy correction per unit cell of the crystal in the kaon condensed phase as
\begin{equation}
    E_{\rm{quant}}=\frac{1}{2\Lambda_3}(I_3^2-\Delta^2).
\label{quant_corr_kcond}
\end{equation}
The first term on the rhs is just the isospin correction, while now there is an additional second term due to the isospin of the kaons. Indeed, since the kaon field enters also in the expression of the isospin moment of inertia $\Lambda_3$, both terms will depend nontrivially on the kaon vev field.

\section{Results}
When the kaon field develops a nonzero vev, apart from the neutron decay and lepton capture processes of \cref{eq:URCA}, additional processes involving kaons may occur:

\begin{equation}
    n\leftrightarrow p + K^-, \qquad l\leftrightarrow K^-+\nu_l
\end{equation}
such that the chemical equilibrium conditions 
\begin{equation}
 \mu_n=\mu_p+\mu_K ,  \qquad  \mu_ l = \mu_ K
 \label{chemeqcond}
\end{equation}
are satisfied. These are the extension of \cref{betacond} to the condensate phase.

The total energy within the unit cell may be obtained as the sum of the baryon, lepton and kaon contributions:
\begin{equation}
\begin{split}
    E = E_{\rm class} +& E_{\rm iso}(\gamma,\tilde{\phi}) + E_K(\mu_e, \tilde{\phi})+\\
   +& E_{e}(\mu_e)+\Theta(\mu_e^2-m_\mu^2)E_{\mu}(\mu_e)
    \label{eq:Etot}
\end{split}
\end{equation}
The kaon contribution is the effective potential energy
\begin{equation}
    E_K(\mu_e, \tilde{\phi}) = V_K-\frac{\Delta^2}{2\Lambda_3},
\end{equation}
which depends on the condensate $\tilde{\phi}$ and on the lepton chemical potential through the explicit dependence on $\mu_K$ of both $V_K$ and $\Delta$, and $\mu_K=\mu_e$ due to the equilibrium conditions \eqref{chemeqcond}. Therefore, the energy of the full system depends on the proton fraction, the kaon vev field and the electron chemical potential. Their respective values can be obtained, for fixed $n_B$ (or equivalently, fixed $L$) by minimizing the free energy
\begin{equation}
    \Omega = E -\mu_e (N_e +\Theta(\mu_e^2-m_\mu^2)N_\mu-\gamma B)
\end{equation}
 with respect to $\gamma$, $\tilde{\phi}$ and $\mu_e$, i.e.
\begin{equation}
    \pdv{\Omega}{\gamma}\eval_{n_B}\!\!\!(\gamma,\tilde{\phi}, \mu_e)=\pdv{\Omega}{\tilde{\phi}}\eval_{n_B}\!\!\!(\gamma,\tilde{\phi}, \mu_e) = \pdv{\Omega}{\mu_e}\eval_{n_B}\!\!\!(\gamma,\tilde{\phi}, \mu_e)=0.
    \label{minimcond}
\end{equation}
The first equation imposes the expected condition $\mu_e = \mu_I = 2\hbar^2(1-2\gamma)/\Lambda_3$. Then, after substituting into the other two conditions we get:
\begin{align}
    \gamma n_B - &\frac{(\mu^2_I - m^2_{e})^{3/2} + (\mu^2_I - m^2_{\mu})^{3/2}}{3\pi^2\hbar^3} + \frac{n_B}{4}\pdv{E_K}{\mu_e}\eval_{\mu_e = \mu_I} = 0, \label{cond1} \\
    &\pdv{V_K}{\tilde{\phi}} - \frac{\Delta}{\Lambda_3}\pdv{\Delta}{\tilde{\phi}} + \pdv{\Lambda_3}{\tilde{\phi}}\left(\frac{\Delta^2}{2\Lambda_3^2} - \frac{\mu_I^2}{2\hbar^2}\right) = 0,
    \label{cond2}
\end{align}
which are precisely the charge neutrality condition, and the minimization of the grand canonical potential with respect to the kaon field. We note here that we drop the ultrarrelativistic consideration for electrons since the appearance of kaons may decrease hugely the electron fraction.
By solving the system of equations \ref{cond1} and\eqref{cond2} for $\gamma$ and $\tilde{\phi}$ we obtain all the needed information for the new kaon condensed phase. Then we may compare the particle fractions and energies between both phases, which we will call $npe\mu$ and $npe\mu\overline{K}$.

Before solving the full system for different values of the lattice length $L$, we may try to obtain the value of the length at which kaons condense, $L_{\rm cond}$. This value is indeed important since it will determine wether or not a condensate of kaons will appear at some point in the interior of NS
This is accomplished with the same system of \cref{cond1,cond2} by factoring the $\sin\tilde{\phi}$ from the second equation and setting $\tilde{\phi} = 0$. Then we may see the system as a pair of equations to obtain the values of $\gamma_{\text{cond}}$ and $L_{\text{cond}}$, the values of the proton fraction and the length parameter for which the kaons condense.

We show in the table below the density at which kaons condense for different values of the parameters as well as the values of some nuclear observables they yield. All the values are given in units of MeV or fm, respectively.
\begin{table*}[h!]
    \centering
    \begin{tabular}{|c|c|c|c|c|c|c|c|c|}
    \hline
         label & $f_{\pi}$ & $e$ & $\lambda^2$ & $E_0$ & $n_0$ & $S_0$ & $L_{\text{sym}}$ & $n_{\text{cond}}/n_0$  \\
         \hline
         set 1 & 133.71 & 5.72 & 5 & 920 & 0.165 & 23.5 & 29.1 & 2.3\\ \hline
         set 2 & 138.11 & 6.34 & 5.78 & 915 & 0.175 & 24.5 & 28.3 & 2.2\\ \hline
         set 3 & 120.96 & 5.64 & 2.68 & 783 & 0.175 & 28.7 & 38.7 & 1.6\\ \hline
         set 4 & 139.26 & 5.61 & 2.74 & 912 & 0.22 & 28.6 & 38.9 & 1.6\\ \hline
        \end{tabular}
    \caption{Sets of parameter values and observables at nuclear saturation}
    \label{tab:Condensation}
\end{table*}

Parameter sets 1 and 2 are chosen so that the energy per baryon and baryon density at saturation are fitted to experimental values, whereas the sets 3 and 4 correctly fit the symmetry energy and slope at saturation.

\begin{figure}
    \centering
    \includegraphics[scale=0.45]{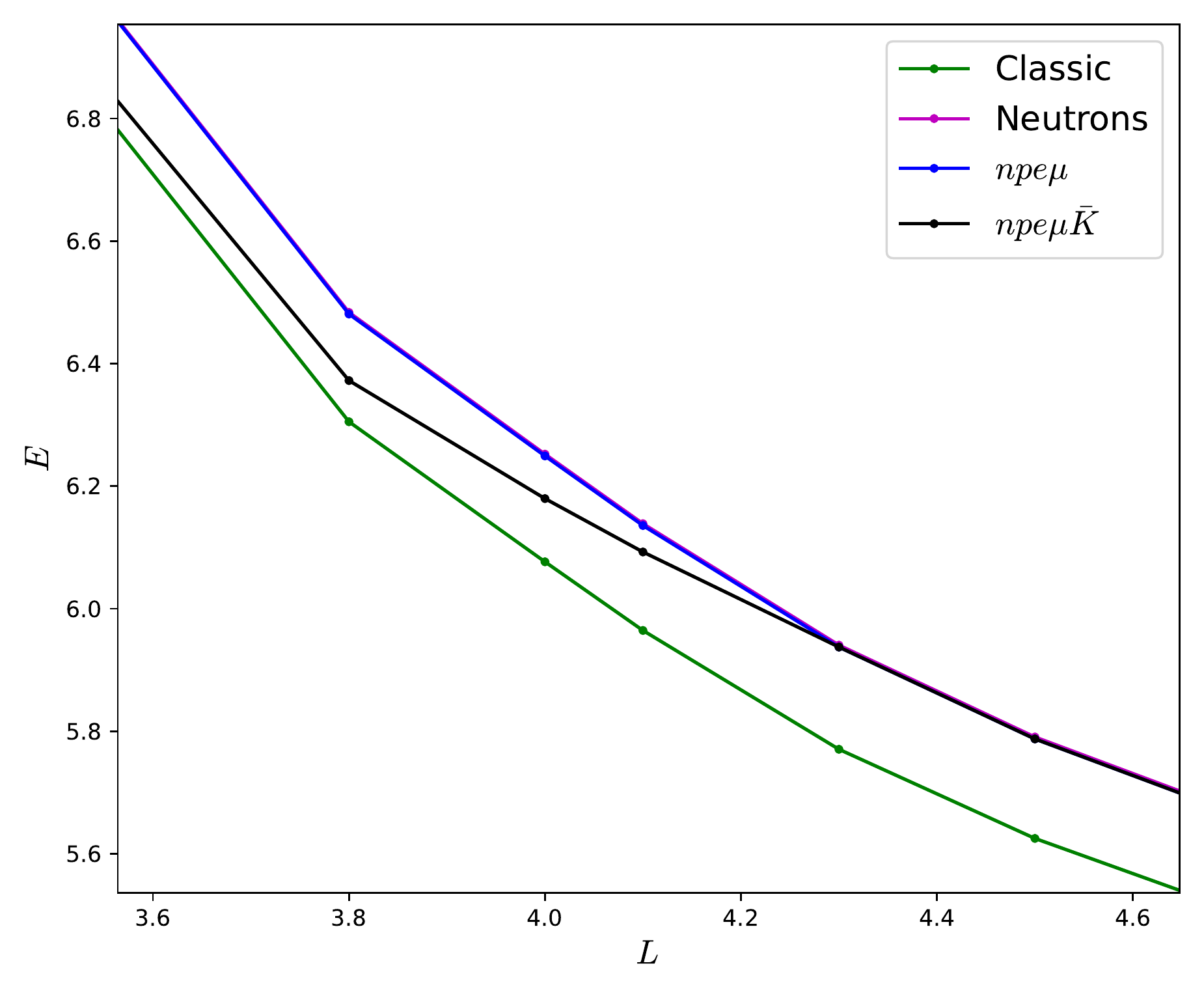}
    \caption{Energy vs lattice length in dimensionless Skyrme units, for the set 1 of parameters. The energy is shown for the classical crystal without isopin contributions (green), isospin asymmetric ($npe\mu$) matter with (black) and without (blue) kaons. We also plot the completely asymmetric neutron matter (magenta) which lies slightly above the blue curve.}
    \label{fig:EvLkaons}
\end{figure}

\begin{figure}
    \centering
    \includegraphics[scale=0.45]{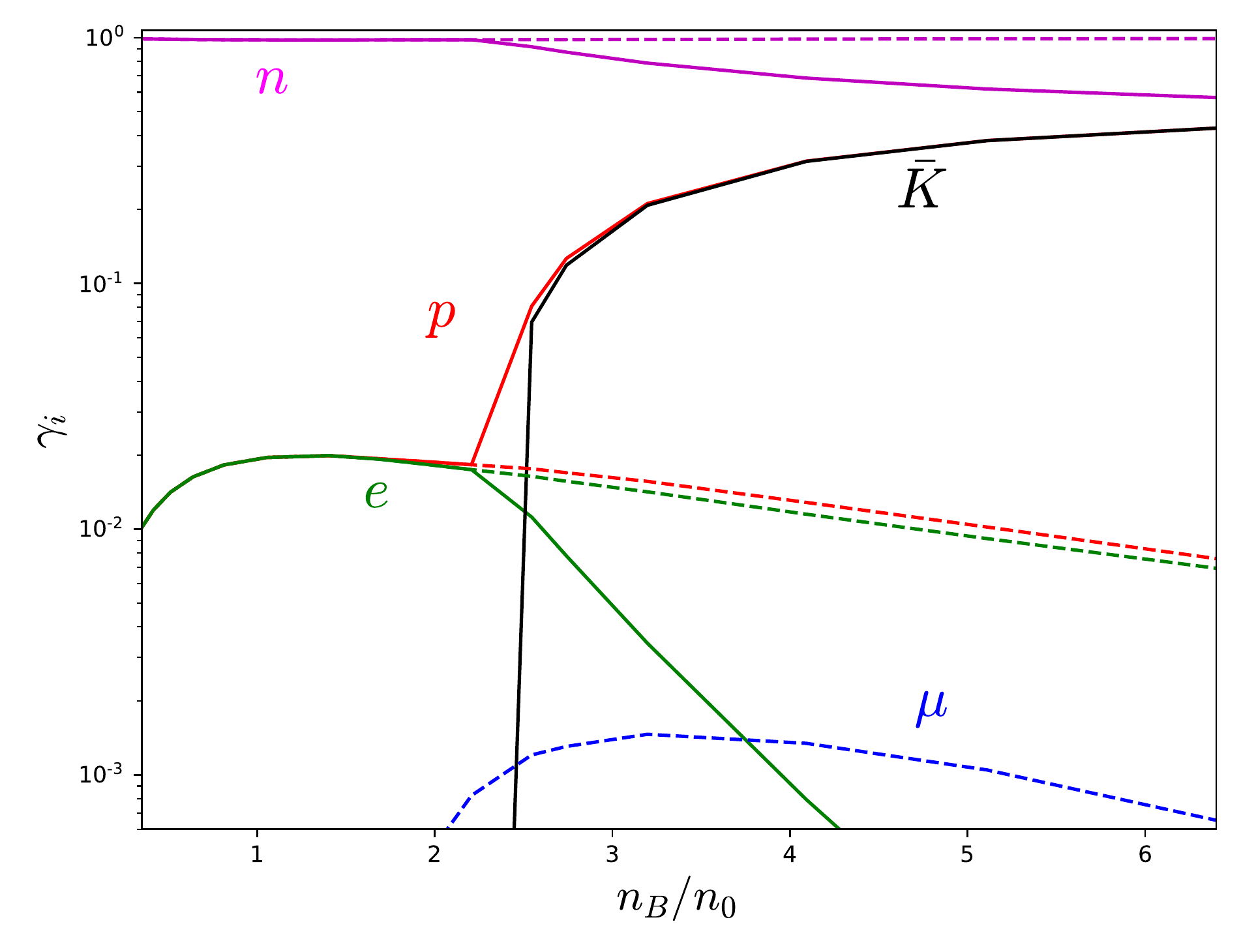}
    \caption{Particle fractions as a function of baryon density for the set 1 of parameters, both with (solid lines) and without (discontinuous lines) kaon condensate. For the case with kaon condensate, the contribution of muons is negligible. }
    \label{fig:Fractions}
\end{figure}

\begin{figure}
    \centering
    \includegraphics[scale=0.5]{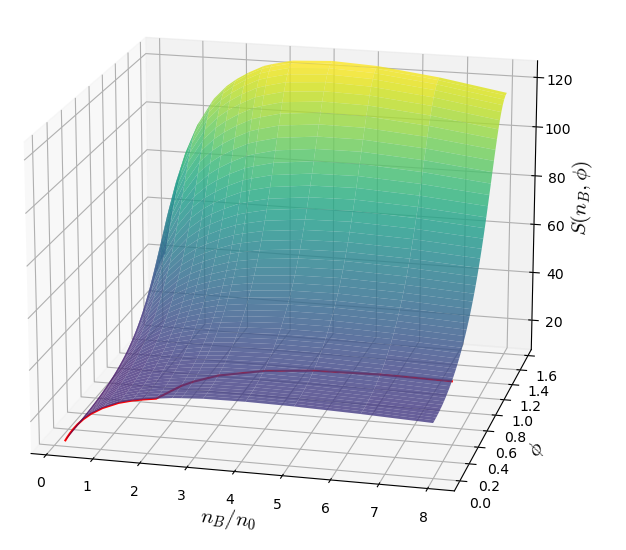}
    \caption{Symmetry energy of nuclear matter as a function of baryon density and kaon $vev$, for the parameter set 1. The surface is obtained by treating  $n_B$ and $\phi$ as independent variables, whereas the red curve corresponds to the energy-minimizing solution for each $L$, i.e., $n_B$.  }
    \label{fig:kaon_fractions_2460}
\end{figure}

In \cref{fig:EvLkaons}, we show the $E(L)$ curves both without and with kaon condensation, in dimensionless Skyrme units. It is clearly visible that for sufficiently small $L$ a nonzero kaon condensate is preferred. In \cref{fig:Fractions} we show the resulting particle fractions. In \cref{fig:kaon_fractions_2460} we plot the symmetry energy as a function of both $n_B$ and the kaon condensate $\phi$. 


\section{Neutron stars with kaon condensed cores within the Skyrme model.}
In this section we briefly recapitulate how to obtain the EoS from the Skyrme crystal solution. Then we calculate the full $npe\mu\bar{K}$ matter EoS, and finally solve the TOV system and compare the NS properties with and without kaons.

\subsection{The Skyrme crystal EoS for (a-)symmetric nuclear matter.}

The energy per baryon as a function of the lattice length, $E(L)$, has a minimum at a certain value $L_0$.
The density $n_0 = (1/2)L_0^{-3}$ at which this minimum is achieved is the so called nuclear saturation density, which has been experimentally found to be $n_0 \simeq 0.16$ fm$^{-3}$. Further, the energy per baryon at nuclear saturation is $E_0 = 923\, {\rm MeV}$.
On the other hand, for the crystal solutions one can define the energy density $\rho$, pressure $p$ and baryon density $n$ as
\begin{align}
    \rho &= \frac{E}{V} = \frac{E_{\text{cell}}}{V_{\text{cell}}}, \\[2mm]
    p &= - \frac{\partial E}{\partial V}= -\frac{\partial E_{\text{cell}}}{\partial V_{\text{cell}}} , \\[2mm]
    n_B &= \frac{B}{V}= \frac{B_{\text{cell}}}{V_{\text{cell}}} = \frac{1}{2L^3}.
\end{align}
We can understand the cell length parameter $L$ as labelling the different energy-minimizing configurations at different densities. Therefore, the three quantities above are related as functions of $L$. This relation is precisely the equation of state (EoS) for skyrmion crystals. It has been recently shown in \cite{Adam:2020yfv} that this EoS stiffens in the generalized Skyrme model, i.e. when the sextic term in \eqref{Energy} is included. This stiffening significantly rises the maximal NS masses that can be reached, which is a strong motivation for the inclusion of the sextic term, as it is necessary in order to reach the mass range of massive pulsars, around $2-2.5M_\odot$ according to recent observations.

The global energy minimum of the crystal is reached at $L=L_0$, at which skyrmionic matter remains in equilibrium, \textit{i.e.} at zero pressure. For $L<L_0$ the crystal is squeezed, which translates into larger values of pressure and density. In the opposite region, $L > L_0$, the matter content inside the unit cell spreads and we enter the unstable branch. Indeed, the pressure in this region is negative, hence we conclude that this (low-density) regime is not well described by the crystal solution. We actually expect that matter inside the unit cell will rearrange in a kind of inhomogeneous central lump surrounded by vacuum \cite{Adam:2021gbm}. More exotic configurations presenting non-homogeneous structures have also been constructed in the Skyrme model, \cite{Park:2019bmi,Canfora:2020kyj}. However, the physical relevance of these configurations as true energy minimizers in the model still remains unclear.

\begin{figure}
    \centering
    \includegraphics[scale=0.32]{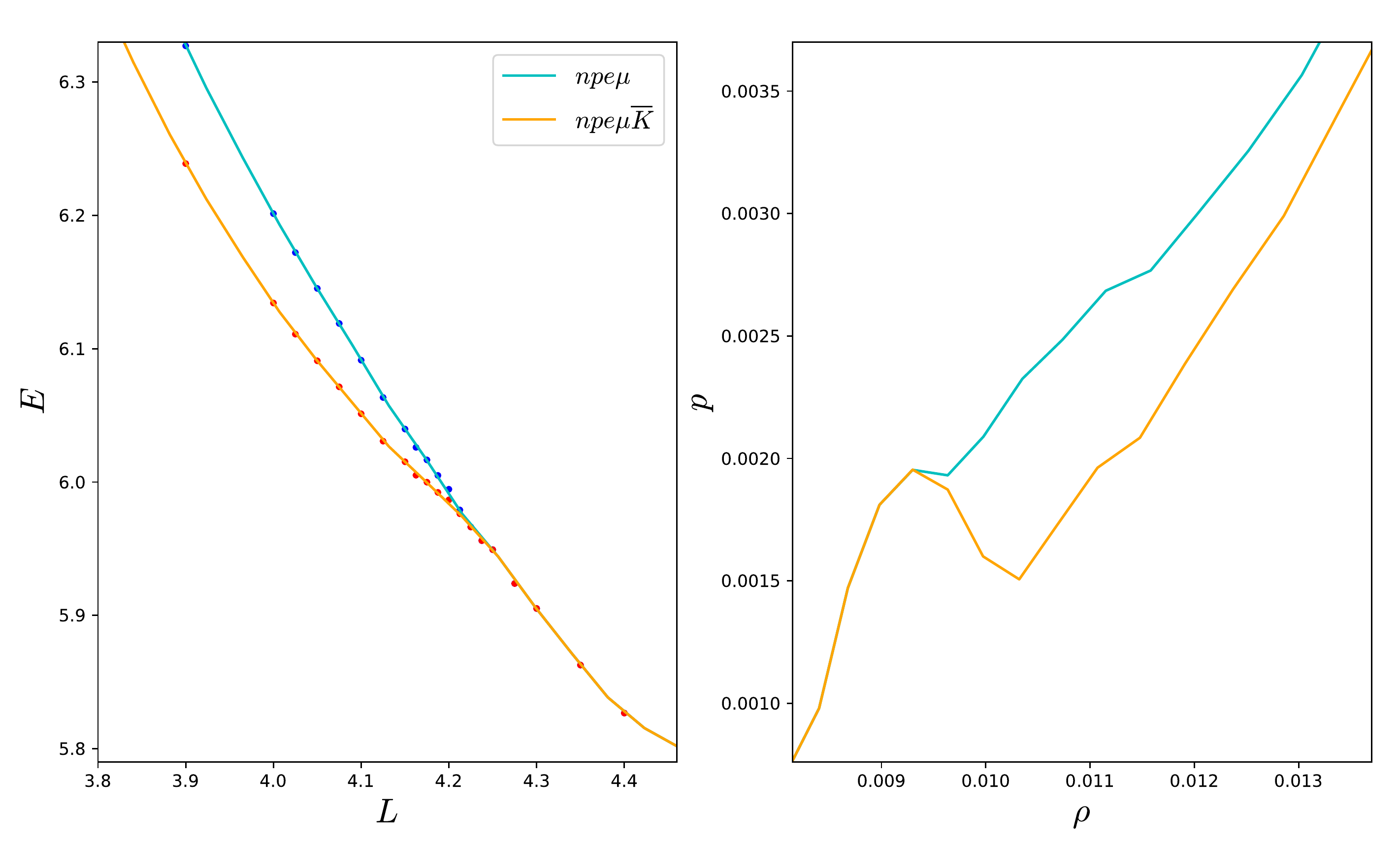}
    \caption{Left plot: energy against the side length of the crystal, calculated with more points near the condensation values for both branches and their interpolations. Right plot: pressure against the energy density (both in Skyrme units) from which we conclude that there is a first order phase transition.}
    \label{Fig.FOT}
\end{figure}

Hence, the main ingredient to obtain the EoS for the Skyrme crystal is the energy dependence on the unit cell size. The $npe\mu$ matter case is easy to obtain using \cref{npemu_matter} for different values of $L$ after solving the $\beta-$equilibrium and charge neutrality conditions for $\gamma$. However, once we include kaons, the change in the energy curve \cref{fig:EvLkaons} may lead to a first or second order phase transition. To distinguish the order of the phase transition in our case, we need to know accurately the pressure near the condensation point. Therefore, we computed more points for the energy near the condensation value with higher accuracies, and we obtained the pressure using a numerical derivative. We conclude that the kaon condensation produces a first order phase transition for our choices of parameters in the Skyrme model. This can be seen in the right plot of \cref{Fig.FOT}, where we show the EoS for our best accuracy and, clearly, there is a non-physical region which must be bridged by a first order phase transition.

\subsection{Maxwell construction versus Gibbs construction}

The Maxwell construction (MC) is typically used to obtain a physical equation of state when a first order transition is present. Indeed, the MC has been already studied in the Skyrme crystals context to describe the transition between crytals with different symmetries \cite{Adam:2021gbm}. This construction is based on a mixed phase of constant pressure which connects the two solutions. However, the MC is only correct when there is a single conserved charge (in this case, the baryon number) for which the associated chemical potential is enforced to be common for both phases in the mixed phase \cite{Glendenning:1992vb}. If, instead, an additional charge is conserved, like the electric charge in the case of $npe\mu$ matter, the Gibbs conditions for the phase equilibrium,
\begin{equation}
    p^{\rm I} = p^{\rm II}, \hspace{2mm} \mu^{\rm I}_i = \mu^{\rm II}_i, \hspace{1.5mm} i = B, q
    \label{Gibbs_conds}
\end{equation}
cannot be both satisfied in a standard MC. In the last expression $\mu_B$ and $\mu_q$ represent the chemical potentials associated to the conserved baryon and electric charges, respectively. Instead, one should perform a Gibbs construction (GC) \cite{Glendenning:1992vb,Glendenning:1997ak}. Indeed, the GC has also been proven useful in the context of a hadron-to-quark phase transition inside NS \cite{Bhattacharyya:2009fg}.

We may write the chemical potential of each particle species as a linear combination of the chemical potentials associated to the conserved charges of our system:
\begin{equation}
    \mu_i = B_i\mu_B + q_i \mu_q,
\end{equation}
where $B_i$ and $q_i$ are the baryon number and electric charge of the particle species $i$.
Then we might identify the baryon and electric charge chemical potentials with the neutron and electron chemical potentials respectively.
The main difference between MC and GC is that, in the mixed phase, the first one imposes charge neutrality locally, \textit{i.e.} both phases are neutral independently, however in the GC it is imposed globally in the mixed phase. Considering a volume fraction $\chi$ of the kaon condensed phase, charge neutrality is imposed in the GC as:
\begin{equation}
    n^{MP}_q = (1-\chi)n^{\rm I}_q + \chi n^{\rm II}_q = 0.
    \label{Charge_neutral_MP}
\end{equation}
The mixed phase in the GC is calculated by identifying first the contributions to the pressure and charge densities in each phase separately. Then we have to solve the system of equations composed by \cref{Gibbs_conds,Charge_neutral_MP,cond2}. We use the unit cell length parameter of the first ($npe\mu$) phase $L_{\rm I}$ as the variable defining our position in the phase diagram, then the unknowns are the length in the second ($npe\mu\bar{K}$) phase $L_{\rm II}$, the proton fractions $\gamma_{\rm I}$, $\gamma_{\rm II}$, the kaon field $\tilde{\phi}$ and the volume fraction $\chi$.
\begin{figure}
    \centering
    \includegraphics[scale=0.35]{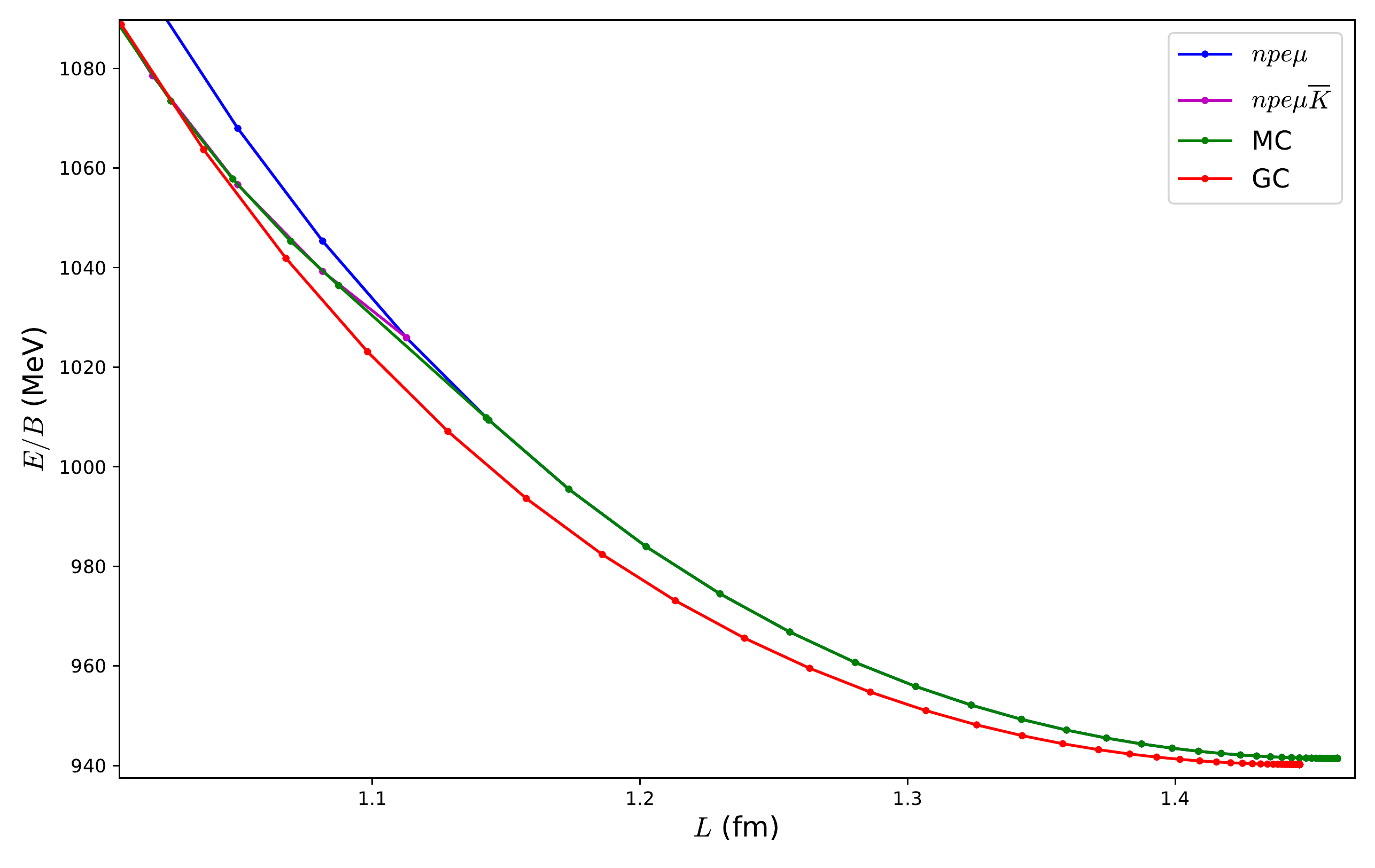}
    \caption{$E(L)$ curves for the two phases. The different slopes at the point of phase separation indicate a first-order phase transition. We also show the curves resulting from a Maxwell construction (MC) and a Gibbs construction (GC). }
    \label{Comparison_MCGC}
\end{figure}

We remark that we assumed in our calculations of the kaon condensate in section III that the backreaction of the condensate on the crystal is negligible, such that our two phases are always considered in the same classical crystal background, and the energies per baryon of the two phases are compared for the same length $L$. As a result, we always should have $L_{\rm I} = L_{\rm II}$ and, consequently, $n_{B,{\rm I}} = n_{B,{\rm II}}$ by construction. On the other hand, the relation between $L$ and the thermodynamical variables $p$, $\mu_i$ and $\phi$ used in \cref{Gibbs_conds,Charge_neutral_MP,cond2} is quite nontrivial in both phases. We, therefore, treat $L_{\rm II}$ as an independent variable in our numerical calculations. We find that always $L_{\rm I} = L_{\rm II}$ within our numerical precision, which provides us with an additional consistency check both for our numerics and for the thermodynamical transformations we used. 

\begin{figure}
    \centering
    \includegraphics[scale=0.37]{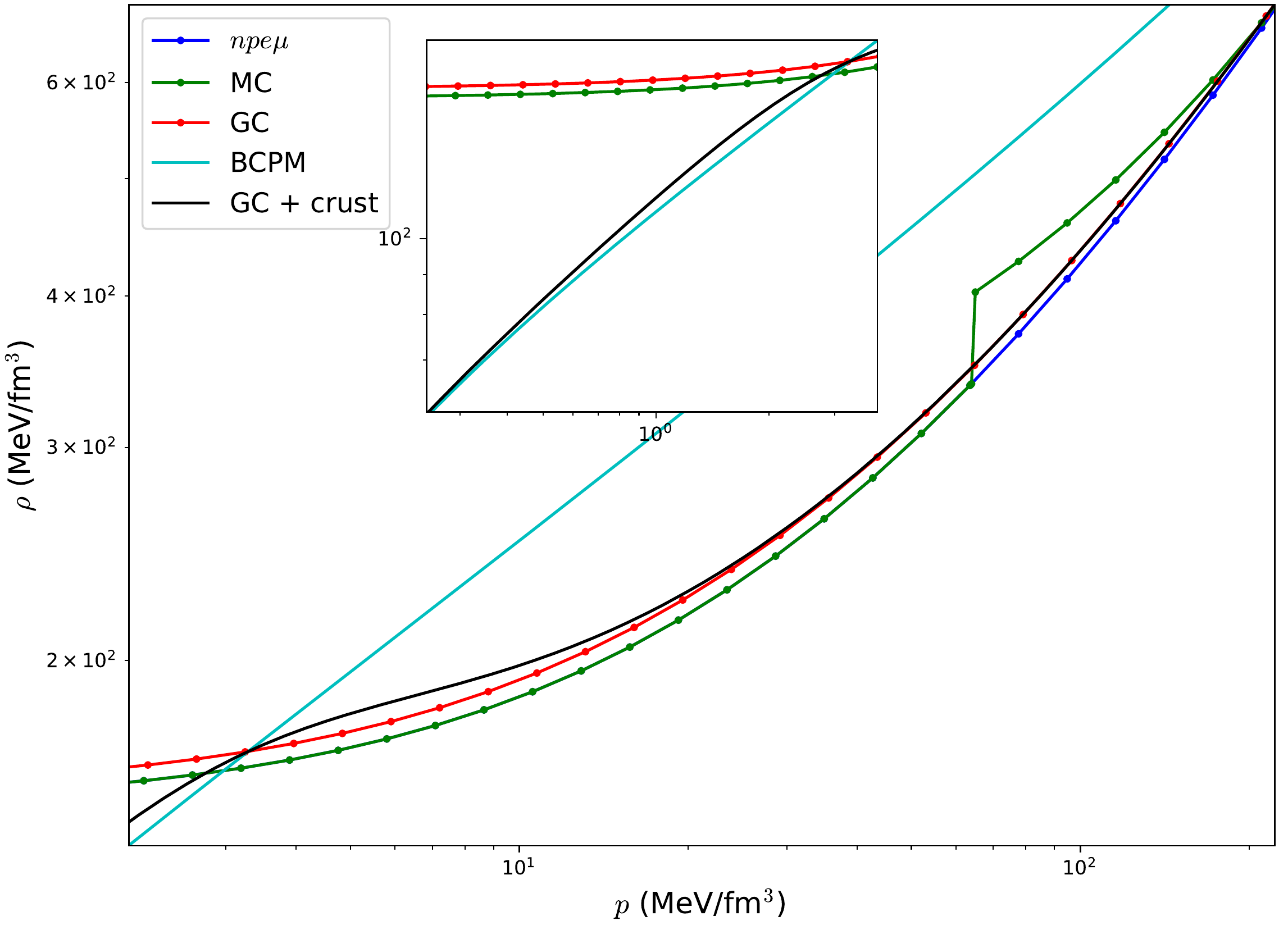}
    \caption{EoS for the three different cases that we have constructed. The jump in the MC due to the first order transition and the different behaviour of the GC at low densities are clearly visible. We also show the standard nuclear physics EoS of \cite{Sharma:2015bna} (BCPM) and a hybrid EoS obtained by joining the BCPM EoS at low pressure with the GC EoS at high pressure.}
    \label{EOS}
\end{figure}

We show our results in \cref{Comparison_MCGC} and in \cref{EOS}.
We find that the mixed phase of the GC, and hence the values at which the kaon field becomes non-zero, starts at a smaller density than the value obtained in \cref{tab:Condensation}. This is also found in \cite{Glendenning:1997ak}, for which the GC mixed phase extends to a larger region than the one obtained from the MC, because the mixed phase in the GC no longer is for constant pressure. In our case,  even the minimum of $E(L)$ is shifted to slightly lower values, see the insert in \cref{Comparison_MCGC} and, hence, the use of the GC affects the low density regime of the EoS, as can be seen in \cref{EOS}. 

\begin{figure}
    \centering
    \includegraphics[scale=0.45]{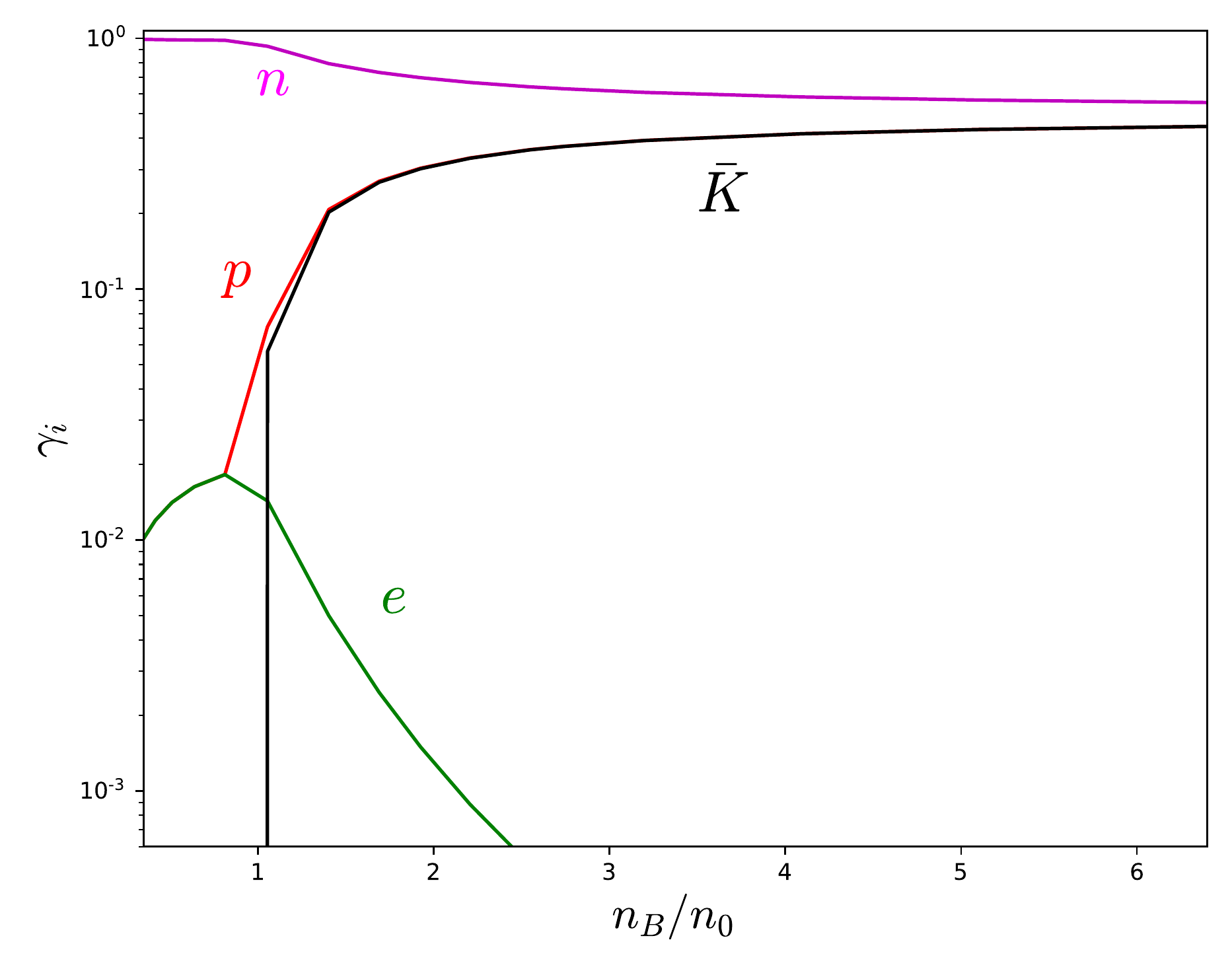}
    \caption{Particle fractions for the GC. The main difference with respect to \cref{fig:Fractions} is the earlier appearance of kaons.}
    \label{FractionsGC}
\end{figure}

We may also calculate the particle fractions in the mixed phase of the GC using an expression equivalent to \cref{Charge_neutral_MP} for each particle. We show the new particle fractions in \cref{FractionsGC}. Besides, during the mixed phase, we find that there are more protons in the second phase than in the first one. However, the presence of more kaons than protons in the second phase results in a partial negative charge density. That negative charge is compensated by the overall positive charge density of the first phase. In both phases the number of electrons is much less than that of protons and kaons.

\subsection{The TOV system and NS properties}

In order to calculate the mass and radius for a non-rotating NS we have to solve the standard TOV (Tolman-Oppenheimer-Volkoff) system of ODEs. It is obtained inserting a spherically symmetric ansatz of the spacetime metric,
\begin{equation}
    ds^2 = -A(r)dt^2 + B(r)dr^2 + r^2(d\theta^2 + \sin^2\theta d\varphi^2),
\end{equation}
in the Einstein equations,
\begin{equation}
    R_{\mu\nu} - \frac{1}{2}g_{\mu\nu}R = 8\pi G T_{\mu\nu}.
\end{equation}
To describe matter inside the star, in the right-hand side of the equation, we use the stress-energy tensor of a perfect fluid,
\begin{equation}
    T_{\mu\nu} = (\rho + p)u_{\mu}u_{\nu} + p g_{\mu\nu},
\end{equation}
where the pressure $p$ and the energy density $\rho$ are not independent but related by the EoS. Hence the EoS describes the nuclear interactions inside the NS and different EoS lead to different observables.

The resulting TOV system involves 3 differential equations for $A, B$ and $p$, which must be solved for a given value of the pressure in the centre of the NS ($p(r = 0) = p_0$) until the condition $p(r = R) = 0$ is achieved.

We use a 4$^{\rm th}$ order Runge-Kutta method of step $\Delta r = 1$ m to solve the system and to obtain the main observables from the solutions. The radial point at which the pressure vanishes defines the radius of the NS, and the mass $M$ is obtained from the Schwarzschild metric definition outside the star,
\begin{equation}
    B(r = R) = \frac{1}{(1 - \frac{2GM}{R})}.
\end{equation}

The results of this section are plotted in \cref{MRcurves} for the 4 sets of parameters. We compare the results between the MC and GC as well as with the EoS without kaons. The first observation is that the addition of kaons to the EoS agrees with the expectation, reducing the achievable maximum mass. This represents the so-called \textit{hyperon puzzle} in which the appearance of new strange degrees of freedom softens the EoS such that it may not lead to sufficiently massive NS ($\sim 2M_{\odot}$). As can be seen, this is not the case in the generalized Skyrme model since we may obtain very high masses easily due to the contribution of the sextic term. Furthermore the radii of NS are also reduced, which benefits our concrete model since the radii for skyrmion crystals are in some cases too large.

The main difference between the two different constructions is that the MC starts at a given density, hence it deviates from the $npe\mu$ EoS at a certain mass. On the other hand, since the GC changes the location of the minimum, it leads to different results also in the low mass region. However, both constructions practically merge in the high masses region, in which they follow the same $npe\mu\overline{K}$ EoS.

As already explained, the thermodynamically stable region of the $E(L)$ curves and the corresponding EoS based on the Skyrme crystal is $L\le L_0$ or, equivalently, $n_B \ge n_0$.
As a consequence, NS based on the Skyrme crystal have $n_B = n_0$ at the NS surface or, in other words, Skyrme crystal NS have no crust.
In the right panel of \cref{MRcurves}, therefore, we show the result of adding a crust to the NS by joining the GC equations of state of the Skyrme crystal with a standard nuclear physics EoS for low densities (the corresponding EoS are shown in \cref{EOS}). Concretely, we use the BCPM EoS \cite{Sharma:2015bna} and joint the two EoS at the pressure $p=p_*$ where the two EoS coincide, i.e., $\rho_{\rm BCPM} (p_*)= \rho_{\rm, crystal}(p_*)$, exactly as we did in
\cite{Adam:2020yfv}. In terms of the baryon density, the joining occurs at $n_{B,*} \sim 1.1 n_0$ for the parameter sets 1-3, and for $n_{B,*} \sim 1.2 n_0$ for the set 4. Again as in 
\cite{Adam:2020yfv}, we assume a smooth joining between the two EoS (concretely, described by a quadratic interpolation) in order to avoid an artificial phase transition at $p_*$.
We also plot in \cref{MRcurves} the most likely
mass-radius relations for the NS corresponding to GW170817 \cite{Abbott_2017} and GW190425 \cite{Abbott:2020uma} events (orange and blue regions). The green regions represent the estimations for the mass and radius values of PSR J0740+6620 (top) \cite{Miller:2019cac} and J0030+0451 (bottom) \cite{Riley:2021pdl}. The purple region constraints the mass-radius curves from the statistical analysis done in \cite{Altiparmak:2022bke}. We find that the NS resulting from the addition of a crust to the Skyrme crystal EoS with a nonzero kaon condensate agree very well with these recent constraints. Further, the softening of the EoS due to the presence of kaons and the resulting smaller NS radii are important for this agreement.

\newpage
\begin{onecolumngrid}

\begin{figure*}[h]
   \centering
    \includegraphics[scale=0.7]{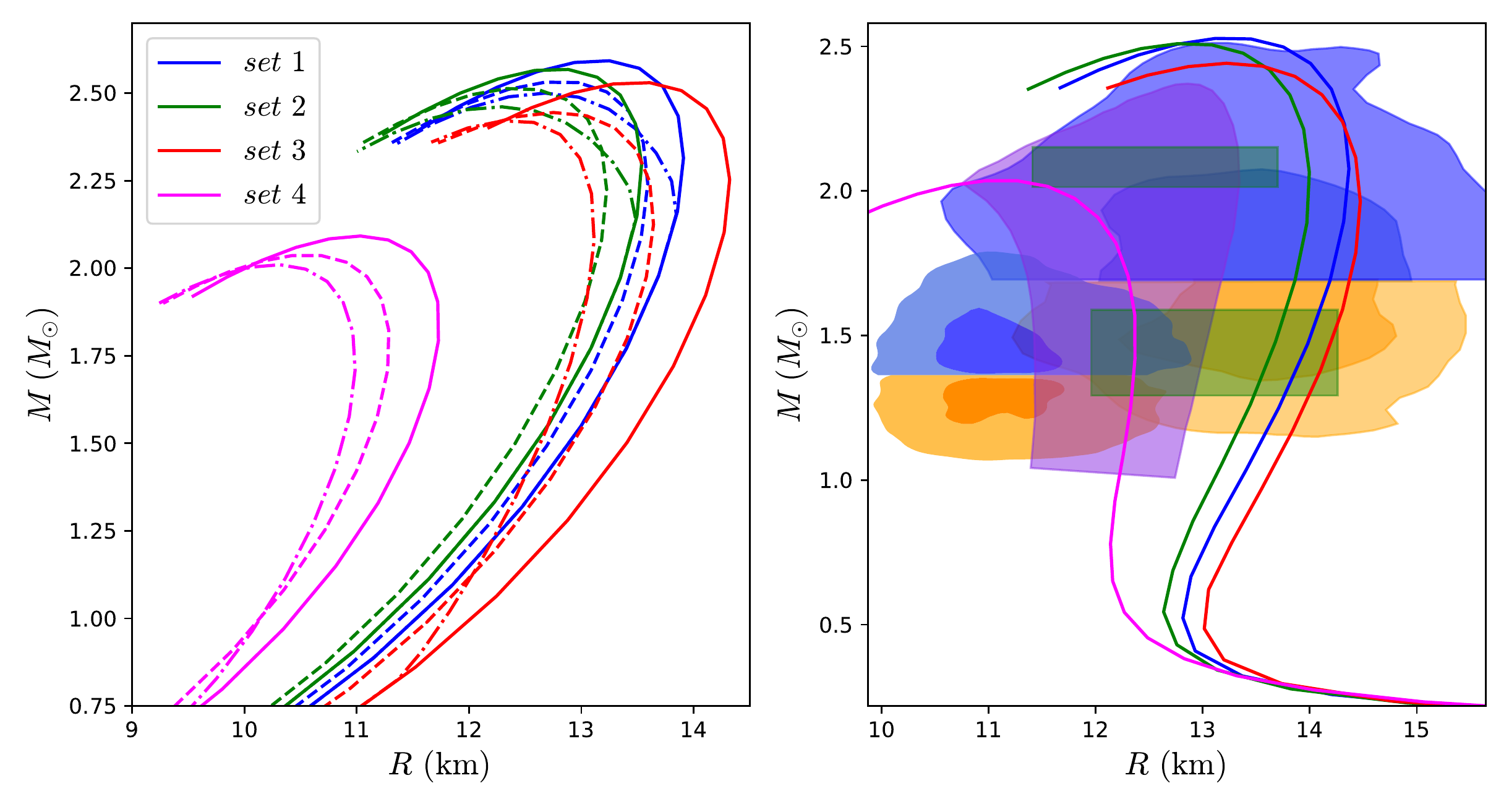}
    \caption{Mass-Radius curves of NS with a kaon condensed core. The different sets of parameters that we consider are shown with different colors. {\em Left panel:}  Solid lines represent $npe\mu$ matter, dashed-dotted lines are obtained with a MC and the dashed with the GC. {\em Right panel:} The effect of adding a standard nuclear physics crust to the Skyrme crystal EoS with kaon condensate obtained from the Gibbs construction (GC). }
    \label{MRcurves}
\end{figure*}
\end{onecolumngrid}
\twocolumngrid

\section{Conclusions}
The quantization of the isospin degrees of freedom in \cite{Adam:2022aes} allowed us to find the dependence of the isospin contribution to the energy (hence the isospin chemical potential) on the lattice length of the skyrmion crystals, which in turn determines the proton (and electron) fraction in $\beta$-equilibrated skyrmion crystals. 
In this paper, we have used this information to precisely determine the critical density at which charged kaons will condense inside neutron stars described by the Skyrme crystal. Although the prediction of the condensation of charged mesons at high (isospin) chemical potential is common in nuclear matter literature, and there have been some partial results within the Skyrme model \cite{Westerberg:1994hu,Park:2009mg}, we have, to our knowledge, 
for the first time provided a framework which allows to precisely calculate the value of the density where kaon condensation sets in. This value turns out to be around twice nuclear saturation for physically relevant sets of parameters, so the prediction from the Skyrme model is that strangeness will be present at the core of neutron stars in the form of an (anti)kaon condensate. Further, we have computed the EoS for skyrmion matter in the condensate phase, which becomes softer due to the additional degrees of freedom. This has appreciable effects on the Mass-Radius curves for physically relevant parameters in the model, reducing the maximum radii by about 0.5-1\,km. We found that a correct treatment of the resulting first-order phase transition between the phase with and without kaon condensate by a Gibbs construction, as originally advocated in \cite{Glendenning:1992vb}, is important for these results.

Despite the prediction of a very early onset of kaon condensation as compared with other nuclear EoS, the maximum mass limit for neutron stars doesn't get significantly reduced, due to the sextic term being dominant at such densities.
The fact that the presence of a kaon condensate does not pose a problem for reaching high masses in the Skyrme model, however, does not imply that the hyperon problem is completely solved in Skyrmion-based EoS. Indeed, in addition to kaons, one should take into account also hyperon degrees of freedom within the Skyrme model. Whilst hyperons can be succesfully described within this model \cite{Klebanov1990}, and even a proposal to study modifications of hyperon properties in dense matter has bee provided in \cite{Hong:2018sqa}, it is not clear for us how to apply these ideas to Skyrmion crystals. A possibility would be to extend the isospin quantization scheme proposed in \cite{Adam:2022aes} to the three flavor case, and quantize the whole $SU(3)$ rotation group. This method, however, is much more technically involved, and relies on the assumption of an approximate $SU(3)$ symmetry of the Hamiltonian, which only holds for sufficiently large densities.

\begin{acknowledgements}
The authors acknowledge financial support from the Ministry of Education, Culture, and Sports, Spain (Grant No. PID2020-119632GB-I00), the Xunta de Galicia (Grant No. INCITE09.296.035PR and Centro singular de investigación de Galicia accreditation 2019-2022), the Spanish Consolider-Ingenio 2010 Programme CPAN (CSD2007-00042), and the European Union ERDF.
AW is supported by the Polish National Science Centre,
grant NCN 2020/39/B/ST2/01553.
AGMC is grateful to the Spanish Ministry of Science, Innovation and Universities, and the European Social Fund for the funding of his predoctoral research activity (\emph{Ayuda para contratos predoctorales para la formaci\'on de doctores} 2019). MHG is also grateful to the Xunta de Galicia (Consellería de Cultura, Educación y Universidad) for the funding of his predoctoral activity through \emph{Programa de ayudas a la etapa predoctoral} 2021.
\end{acknowledgements}

\appendix

\appendix
\section{Derivation of the WZW and sextic terms contribution to $V_K$}
\label{WZW-contribution}
To work with the WZW and sextic terms, it is useful to employ the formalism of Lie algebra-valued differential forms, which we will extensively do in this appendix. Let us first review the basic properties of such objects and establish the notation that we will follow.
A $\mathfrak{g}$-valued differential form $\alpha$ can be written in terms of the Lie algebra generators $T_a$, as
$\alpha=\alpha^a\otimes T_ a$. The exterior derivative is then simply obtained as $d\alpha = d\alpha^ a\otimes T_ a$. Furthermore, the wedge product on $\mathfrak{g}$-valued forms is defined as 
\begin{equation}
    \alpha\wedge \beta=\alpha^a\wedge \beta ^b \otimes T_ aT_b.
\label{Liewedge}
\end{equation}
So that the following useful properties hold:
\begin{align}
    d\alpha\wedge\beta&=(d\alpha)\wedge\beta+(-1)^\abs{\alpha}\alpha\wedge(d\beta)
    \label{prop1}\\
    \Tr{\alpha\wedge\beta}&=(-1)^{\abs{\alpha}\abs{\beta}}\Tr{\beta\wedge\alpha},
    \label{prop2}
\end{align}
where $\abs{\alpha}$ denotes the degree of $\alpha$. Also, by linearity of the trace, both the trace and the exterior derivative commute, i.e. 
\begin{equation}
    d\Tr{\alpha}=\Tr{d\alpha}.
    \label{prop3}
\end{equation}
To alleviate the notation, in the following we will drop the wedge product symbol and denote the product \eqref{Liewedge} simply by $\alpha\beta$. Then, for instance, if $\alpha$ and $\beta$ denote two 1-forms, we have
$
    \Tr{\alpha\beta} = -\Tr{\beta\alpha}$,  $d(\alpha\beta)=d\alpha\beta-\alpha d\beta.
$

Let us now perform the most general chiral transformation to the Skyrme field $U_\pi$, given by $U=g_l U_\pi g_r^\dagger$, with $(g_l,g_r)\in SU(3)_L\times SU(3)_R$ and define the following $\mathfrak{su}(3)$-valued differential forms, 
\begin{equation}
\begin{split}
    V=U^\dagger &dU, \,\, L = U_\pi^\dagger dU_\pi,\,\, \alpha = g_l^\dagger dg_l,\,\beta=g_r dg_r^\dagger.
\end{split}
\end{equation}
By definition, we have the following relation between the forms above:
\begin{equation}
    V=(U_\pi g_r)^\dagger[\alpha+U_\pi(L-\beta)U^\dagger_\pi]U_\pi g_r.
\end{equation}
On the other hand, the WZW action is then given by the pullback of a volume 5-form $\Omega_5$ by an extended Skyrme field $U:D^5\rightarrow SU(3)$ \footnote{The result is of course independent of such extension, because $\pi_4(SU(3))$ vanishes.} integrated over an auxiliar 5-dimensional disk $D$ whose boundary is the spacetime manifold $M$, 

\begin{equation}
      S_{\rm WZW}=-i\frac{N_C}{240\pi^2}\int_DU^*(\Omega_5)
\end{equation}
The form $U^*(\Omega_5)$ can be expressed in terms of $L$ as
\begin{align}
    S_{\rm WZW}(L)&=-i\frac{N_C}{240\pi^2}\int_D\Tr\{V^5\} =\notag\\
    =&-\frac{iN_C}{240\pi^2}\int_D\Tr\{[\alpha+U_\pi(L-\beta)U^\dagger_\pi]^5\} .
\end{align}

Let us denote the 1-form $U_\pi(L-\beta)U^\dagger_\pi$ by $\omega$. The exterior derivative of this form is:
\begin{align}
    d\omega &=dU_\pi(L-\beta)U^\dagger_\pi+U_\pi(dL-d\beta)U^\dagger_\pi-U_\pi(L-\beta)dU^\dagger_\pi=\notag\\
    &=-U^\dagger_\pi(dL+d\beta+L\beta +\beta L)U_\pi,
\end{align}
where we have used the fact that both $\beta$ and $L$ satisfy the Maurer-Cartan equation $d\beta=-\beta^2$.
Moreover, one can straightforwardly see that
\begin{equation}
    \omega^2=U^\dagger_\pi(L-\beta)^2U_\pi=U^\dagger_\pi(-d\beta-dv-\beta v-v\beta)U_\pi=d\omega
\label{relomeg}
\end{equation}
Knowing this, we have
\begin{widetext}
\begin{equation}
\begin{split}
        S_{\rm WZW}(L)=&-i\frac{N_C}{240\pi^2}\int_\mathcal{M}\Tr\{[\alpha+\omega]^5\} \\
        =&S_{\rm WZW}(\alpha)+S_{\rm WZW}(\omega)-\frac{iN_C}{48\pi^2}\int_\mathcal{M}\Tr\{\alpha^4\omega+\omega^4\alpha+\alpha^2\omega^3+\omega^2\alpha^3+\alpha\omega\alpha\omega^2+\omega\alpha\omega\alpha^2\} \\
        =&S_{\rm WZW}(\alpha)+S_{\rm WZW}(\omega)-\frac{iN_C}{48\pi^2}\int_{\partial\mathcal{M}}\Tr\{\omega^3\alpha-\alpha^3\omega-\frac{1}{2}(\alpha\omega)^2\},
\end{split}
\end{equation}
\end{widetext}
where we have used \cref{prop1,prop2,prop3}, the relation \eqref{relomeg} for $\omega$, the M-C equation for $\alpha$ and Stokes' theorem in the last step.
Repeating the same calculation for $S_{\rm WZW}(\omega)$ yields:
\begin{align}
        S_{\rm WZW}(\omega)=&-i\frac{N_C}{240\pi^2}\int_\mathcal{M}\Tr\{[L-\beta]^5\} =\\
        =&-S_{\rm WZW}(\beta)+S_{\rm WZW}(L)-\notag\\
        &-\frac{iN_C}{48\pi^2}\int_{\partial\mathcal{M}}\!\!\Tr\{L^3\beta-\beta ^3L-\frac{1}{2}(L\beta)^2\},\notag
\end{align}
so that 
\begin{align}
        \hspace*{-0.6cm}
        S_{\rm WZW}&(V)=S_{\rm WZW}(L)+S_{\rm WZW}(\alpha)-S_{\rm WZW}(\beta)-\notag
        \\
        -\frac{iN_C}{48\pi^2}&\int_{\partial\mathcal{M}}\Tr\{L^3\beta-\beta ^3L-\frac{1}{2}(L\beta)^2+\omega^3\alpha-\alpha^3\omega-\frac{1}{2}(\alpha\omega)^2\}.\label{TotalWZW}
\end{align}
Eq.\eqref{TotalWZW} shows that a chiral transformation of the $SU(3)$ Skyrme field induces an additional local term in the action due to the nontrivial transformation of the (nonlocal) WZW term.
Furthermore, if we fix the chiral transformation fields to only depend on one spacetime coordinate, $g_{l/r}(x)\equiv g_{l/r}(t)$,  any power of $\alpha$ and $\beta$ will vanish in the local, $4$-dimensional effective term. Taking this into account, we arrive to the final result:
\begin{equation}
    S_{\rm WZW}(V)=S_{\rm WZW}(L)-\frac{iN_C}{48\pi^2}\int_{\partial\mathcal{M}}\Tr\{L^3(\beta+U^\dagger_\pi \alpha U_\pi)\},
\end{equation}
from where \cref{WZWkaoncontrib} is readily obtained.

Let us now turn to the sextic term. The coordinate free version of $\mathcal{L}_6$ is given by
\begin{equation}
    \mathcal{L}_6=\lambda^2\pi^4\mathcal{B}\wedge \star\mathcal{B}
\end{equation}
where $\star$ denotes the Hodge star operator, and $\mathcal{B}$ the 1-form in spacetime whose coordinates in a local chart coincide with the Baryon current, $\mathcal{B}=B_\mu dx^\mu$. We can construct such form as the Hodge dual of the baryon number density three-form,
\begin{align}
    b&=(24\pi^2)^{-1}U^*(\Omega_3)=\frac{1}{24\pi^2}\Tr\{L_\mu L_\nu L_\rho\}dx^\mu\wedge dx^\nu\wedge dx^\rho\notag\\
    &=\frac{1}{3!}b_{\mu\nu\rho}dx^\mu\wedge dx^\nu\wedge dx^\rho,
\end{align} i.e.
\begin{align}
        \mathcal{B}= \star b =& \frac{1}{3!}b^{\nu\rho\sigma}\varepsilon_{\nu\rho\sigma\mu} dx^\mu=\notag\\ =&\frac{1}{24\pi^2}\varepsilon_\mu^{\,\,\,\nu\rho\sigma}\Tr\{L_\nu L_\rho L_\sigma\} = B_\mu dx^\mu.
\end{align}
Thus $\mathcal{L}_6=\lambda^2\pi^4\star b\wedge b$. Expressing the sextic term in such form is most useful for calculating its contribution to the kaon potential employing the formalism of differential forms in the same way as for the WZW term. Indeed, we see that
\begin{equation}
    b=\frac{1}{24\pi^2}\Tr\{L^3\}=\frac{1}{24\pi^2}\Tr\{\omega^3+3w\omega^2\},
\end{equation}
where we used the fact that $w^n=0$ for $n\geq 2$ because the kaon field $\Sigma$ only depends on time. Hence, also $v^n=0$ for $n\geq 2$ and $vw=0$ hold, and, given that $\omega=U_\pi^\dagger(\beta-v)U_\pi$, we may write 
\begin{align}
    b=&\frac{1}{24\pi^2}\Tr\{\beta^3+3(w U_\pi^\dagger(\beta-v)^2 U_\pi-\beta^2 v)\}=\notag\\
    =&\frac{1}{24\pi^2}\Tr\{\beta^3+3\beta^2(U_\pi w U_\pi^\dagger -v)\}.
\end{align}
Thus, the baryon current density in the kaon condensed phase will be modified by $    B^\mu=B^\mu_\pi+C^\mu$, where $B^\mu_\pi$ is the baryon current due to the pionic background and
\begin{equation}
C^\mu= \frac{1}{8\pi^2}\varepsilon^{\mu\nu\rho\sigma}\Tr\{R_\nu R_\rho (U_\pi \Sigma\partial_\sigma\Sigma^\dagger U_\pi^\dagger-\Sigma^\dagger \partial_\sigma \Sigma).\}
\end{equation}
For a time independent pion background and a homogeneous kaon condensate, we have $B_\pi^\mu C_\mu = 0$, and hence the only constribution from the sextic term ($\propto B_\mu B^\mu$) to $V_K$ comes from the additional term:
\begin{align}
    C_\mu C^\mu&=\frac{1}{64\pi^4}\varepsilon^{\mu\nu\rho\sigma}\epsilon_{\mu\alpha\beta}\Tr\{R_\nu RE_\rho \xi_\sigma\}\Tr\{R_\alpha R_\beta \xi_\gamma\}=\notag\\
    &=\frac{-1}{64\pi^4}\varepsilon_{ijk}\varepsilon_{ilm}\Tr\{R_j R_k \xi_0\}Tr\{R_l R_m \xi_0\}
\end{align}
where $\xi_\mu=U_\pi \Sigma\partial_\mu\Sigma^\dagger U_\pi^\dagger-\Sigma^\dagger \partial_\mu \Sigma$.
\section{Explicit expressions}
\label{app2}
We show here the explicit expressions of the (third component of the) inertia tensor and the kaon isospin current, divided into separated contributions coming from the quadratic, quartic, sextic and Wess-Zumino-Witten terms. A superindex labels the origin of each contribution.

\begin{align}
    \Lambda_{33}&\equiv\Lambda_3= \Lambda_3^{(2)}+\Lambda_3^{(4)}+\Lambda_3^{(6)}\\
    \Delta_3&\equiv\Delta = \Delta^{(2)}+\Delta^{(4)}+\Delta^{(6)}+\Delta^{(\rm WZW)}
\end{align}
with
\begin{align}
    \Lambda^{(2)}_3&=2a\Tr\{T_3T_3\}=\\
    &=\frac{\pi_1^2+\pi_2^2}{2}(1+\cos^2\phi)^2+(1+\sigma)\sin^2(2\phi)/4,\notag\\[2mm]
    \Lambda^{(4)}_3&=-4b\Tr\{[T_3,L_k][T_3,L_k]\}=2(1+\cos^2\phi)\times\notag\\
    &\times\big [(1-\pi^2_3)\partial_i\sigma^2+\sigma\pi_3\partial_i\sigma\partial_i\pi_3)+(\sigma\leftrightarrow\pi_3)\big]+\notag\\
    &+\partial_in^2(1+\sigma)/4\sin^2(2\phi),
    \\[2mm]
    \Lambda^{(6)}_3&=-\frac{c}{32\pi^4}\varepsilon^{lmn}\Tr\{T_3L_mL_n\}\varepsilon_{lrs}\Tr\{T_3L_rL_s\}=\notag\\
    =&\frac{\lambda^2 f^2_{\pi} e^4 \mu_K^2}{2} (1+\cos^2(\tilde\phi))^2 (\partial_i\pi_3\partial_j\sigma-\partial_i\sigma\partial_j\pi_3)^2\\[2mm]
    \Delta^{(2)}&=2a\Tr\{T_3L_0\}=\\
    &=-i\mu_K[(\pi_1^2+\pi_2^2)(\cos^4\phi-1)+(1+\sigma)\sin^2(2\phi)/2],\notag\\[2mm]
    \Delta^{(4)}&=-4b\Tr\{[T_a,L_k][T_b,L_k]\}=-2i\mu_K\big[2(1-\cos^4\phi)\times\notag\\
    &\times(\pi_1^2\partial_i\pi_2^2+\pi_2^2\partial_i\pi_1^2-2\pi_1\pi_2\partial_i\pi_1\partial_i\pi_2-\partial_i n^2(\pi_1^2 + \pi^2_2))+\notag\\
    &+\partial_in^2/4(1+\sigma)\sin^2(2\phi)\big],
    \\[2mm]
    \Delta^{(6)}&=-\frac{c}{32\pi^4}\varepsilon^{lmn}\Tr\{T_3L_mL_n\}\varepsilon_{lrs}\Tr\{L_0L_rL_s\}=\notag\\
    &=i\mu_K\lambda^2 f^2_{\pi} e^4 (1+\cos^2\phi)\sin^2\phi (\partial_i\pi_3\partial_j\sigma-\partial_i\sigma\partial_j\pi_3)^2\\[2mm]
    \Delta&^{(\rm WZW)}=-\frac{ N_CB_{\rm cell}}{2}  \sin^{2}{\left(\phi \right)}
\end{align}

\newpage

\begin{thebibliography}{59}%
\makeatletter
\providecommand \@ifxundefined [1]{%
 \@ifx{#1\undefined}
}%
\providecommand \@ifnum [1]{%
 \ifnum #1\expandafter \@firstoftwo
 \else \expandafter \@secondoftwo
 \fi
}%
\providecommand \@ifx [1]{%
 \ifx #1\expandafter \@firstoftwo
 \else \expandafter \@secondoftwo
 \fi
}%
\providecommand \natexlab [1]{#1}%
\providecommand \enquote  [1]{``#1''}%
\providecommand \bibnamefont  [1]{#1}%
\providecommand \bibfnamefont [1]{#1}%
\providecommand \citenamefont [1]{#1}%
\providecommand \href@noop [0]{\@secondoftwo}%
\providecommand \href [0]{\begingroup \@sanitize@url \@href}%
\providecommand \@href[1]{\@@startlink{#1}\@@href}%
\providecommand \@@href[1]{\endgroup#1\@@endlink}%
\providecommand \@sanitize@url [0]{\catcode `\\12\catcode `\$12\catcode
  `\&12\catcode `\#12\catcode `\^12\catcode `\_12\catcode `\%12\relax}%
\providecommand \@@startlink[1]{}%
\providecommand \@@endlink[0]{}%
\providecommand \url  [0]{\begingroup\@sanitize@url \@url }%
\providecommand \@url [1]{\endgroup\@href {#1}{\urlprefix }}%
\providecommand \urlprefix  [0]{URL }%
\providecommand \Eprint [0]{\href }%
\providecommand \doibase [0]{http://dx.doi.org/}%
\providecommand \selectlanguage [0]{\@gobble}%
\providecommand \bibinfo  [0]{\@secondoftwo}%
\providecommand \bibfield  [0]{\@secondoftwo}%
\providecommand \translation [1]{[#1]}%
\providecommand \BibitemOpen [0]{}%
\providecommand \bibitemStop [0]{}%
\providecommand \bibitemNoStop [0]{.\EOS\space}%
\providecommand \EOS [0]{\spacefactor3000\relax}%
\providecommand \BibitemShut  [1]{\csname bibitem#1\endcsname}%
\let\auto@bib@innerbib\@empty
\bibitem [{\citenamefont {Li}\ \emph {et~al.}(2018)\citenamefont {Li},
  \citenamefont {Sedrakian},\ and\ \citenamefont {Weber}}]{Li:2018qaw}%
  \BibitemOpen
  \bibfield  {author} {\bibinfo {author} {\bibfnamefont {J.~J.}\ \bibnamefont
  {Li}}, \bibinfo {author} {\bibfnamefont {A.}~\bibnamefont {Sedrakian}}, \
  and\ \bibinfo {author} {\bibfnamefont {F.}~\bibnamefont {Weber}},\ }\href
  {\doibase 10.1016/j.physletb.2018.06.051} {\bibfield  {journal} {\bibinfo
  {journal} {Phys. Lett. B}\ }\textbf {\bibinfo {volume} {783}},\ \bibinfo
  {pages} {234} (\bibinfo {year} {2018})},\ \Eprint
  {http://arxiv.org/abs/1803.03661} {arXiv:1803.03661 [nucl-th]} \BibitemShut
  {NoStop}%
\bibitem [{\citenamefont {Glendenning}(1985)}]{Glendenning:giant}%
  \BibitemOpen
  \bibfield  {author} {\bibinfo {author} {\bibfnamefont {N.~K.}\ \bibnamefont
  {Glendenning}},\ }\href {\doibase 10.1086/163253} {\bibfield  {journal}
  {\bibinfo  {journal} {Astrophys. J.}\ }\textbf {\bibinfo {volume} {293}},\
  \bibinfo {pages} {470} (\bibinfo {year} {1985})}\BibitemShut {NoStop}%
\bibitem [{\citenamefont {{Hartle}}\ \emph {et~al.}(1975)\citenamefont
  {{Hartle}}, \citenamefont {{Sawyer}},\ and\ \citenamefont
  {{Scalapino}}}]{1975ApJ...199..471H}%
  \BibitemOpen
  \bibfield  {author} {\bibinfo {author} {\bibfnamefont {J.~B.}\ \bibnamefont
  {{Hartle}}}, \bibinfo {author} {\bibfnamefont {R.~F.}\ \bibnamefont
  {{Sawyer}}}, \ and\ \bibinfo {author} {\bibfnamefont {D.~J.}\ \bibnamefont
  {{Scalapino}}},\ }\href {\doibase 10.1086/153713} {\bibfield  {journal}
  {\bibinfo  {journal} {\apj}\ }\textbf {\bibinfo {volume} {199}},\ \bibinfo
  {pages} {471} (\bibinfo {year} {1975})}\BibitemShut {NoStop}%
\bibitem [{\citenamefont {Celenza}\ \emph {et~al.}(1977)\citenamefont
  {Celenza}, \citenamefont {Nutt},\ and\ \citenamefont
  {Shakin}}]{CELENZA197723}%
  \BibitemOpen
  \bibfield  {author} {\bibinfo {author} {\bibfnamefont {L.~S.}\ \bibnamefont
  {Celenza}}, \bibinfo {author} {\bibfnamefont {W.}~\bibnamefont {Nutt}}, \
  and\ \bibinfo {author} {\bibfnamefont {C.}~\bibnamefont {Shakin}},\ }\href
  {\doibase https://doi.org/10.1016/0370-2693(77)90053-3} {\bibfield  {journal}
  {\bibinfo  {journal} {Physics Letters B}\ }\textbf {\bibinfo {volume} {72}},\
  \bibinfo {pages} {23} (\bibinfo {year} {1977})}\BibitemShut {NoStop}%
\bibitem [{\citenamefont {Kaplan}\ and\ \citenamefont
  {Nelson}(1988)}]{Kaplan:1987sc}%
  \BibitemOpen
  \bibfield  {author} {\bibinfo {author} {\bibfnamefont {D.~B.}\ \bibnamefont
  {Kaplan}}\ and\ \bibinfo {author} {\bibfnamefont {A.~E.}\ \bibnamefont
  {Nelson}},\ }\href {\doibase 10.1016/0375-9474(88)90442-3} {\bibfield
  {journal} {\bibinfo  {journal} {Nucl. Phys. A}\ }\textbf {\bibinfo {volume}
  {479}},\ \bibinfo {pages} {273c} (\bibinfo {year} {1988})}\BibitemShut
  {NoStop}%
\bibitem [{\citenamefont {Glendenning}\ and\ \citenamefont
  {Schaffner-Bielich}(1999)}]{Glendenning:1997ak}%
  \BibitemOpen
  \bibfield  {author} {\bibinfo {author} {\bibfnamefont {N.~K.}\ \bibnamefont
  {Glendenning}}\ and\ \bibinfo {author} {\bibfnamefont {J.}~\bibnamefont
  {Schaffner-Bielich}},\ }\href {\doibase 10.1103/PhysRevC.60.025803}
  {\bibfield  {journal} {\bibinfo  {journal} {Phys. Rev. C}\ }\textbf {\bibinfo
  {volume} {60}},\ \bibinfo {pages} {025803} (\bibinfo {year} {1999})},\
  \Eprint {http://arxiv.org/abs/astro-ph/9810290} {arXiv:astro-ph/9810290}
  \BibitemShut {NoStop}%
\bibitem [{\citenamefont {Pal}\ \emph {et~al.}(2000)\citenamefont {Pal},
  \citenamefont {Bandyopadhyay},\ and\ \citenamefont {Greiner}}]{Pal:2000pb}%
  \BibitemOpen
  \bibfield  {author} {\bibinfo {author} {\bibfnamefont {S.}~\bibnamefont
  {Pal}}, \bibinfo {author} {\bibfnamefont {D.}~\bibnamefont {Bandyopadhyay}},
  \ and\ \bibinfo {author} {\bibfnamefont {W.}~\bibnamefont {Greiner}},\ }\href
  {\doibase 10.1016/S0375-9474(00)00175-5} {\bibfield  {journal} {\bibinfo
  {journal} {Nucl. Phys. A}\ }\textbf {\bibinfo {volume} {674}},\ \bibinfo
  {pages} {553} (\bibinfo {year} {2000})},\ \Eprint
  {http://arxiv.org/abs/astro-ph/0001039} {arXiv:astro-ph/0001039} \BibitemShut
  {NoStop}%
\bibitem [{\citenamefont {Heiselberg}\ \emph {et~al.}(1993)\citenamefont
  {Heiselberg}, \citenamefont {Pethick},\ and\ \citenamefont
  {Staubo}}]{PhysRevLett.70.1355}%
  \BibitemOpen
  \bibfield  {author} {\bibinfo {author} {\bibfnamefont {H.}~\bibnamefont
  {Heiselberg}}, \bibinfo {author} {\bibfnamefont {C.~J.}\ \bibnamefont
  {Pethick}}, \ and\ \bibinfo {author} {\bibfnamefont {E.~F.}\ \bibnamefont
  {Staubo}},\ }\href {\doibase 10.1103/PhysRevLett.70.1355} {\bibfield
  {journal} {\bibinfo  {journal} {Phys. Rev. Lett.}\ }\textbf {\bibinfo
  {volume} {70}},\ \bibinfo {pages} {1355} (\bibinfo {year}
  {1993})}\BibitemShut {NoStop}%
\bibitem [{\citenamefont {Benvenuto}\ and\ \citenamefont
  {Lugones}(1999)}]{Benvenuto:1999uk}%
  \BibitemOpen
  \bibfield  {author} {\bibinfo {author} {\bibfnamefont {O.~G.}\ \bibnamefont
  {Benvenuto}}\ and\ \bibinfo {author} {\bibfnamefont {G.}~\bibnamefont
  {Lugones}},\ }\href {\doibase 10.1046/j.1365-8711.1999.02458.x} {\bibfield
  {journal} {\bibinfo  {journal} {Mon. Not. Roy. Astron. Soc.}\ }\textbf
  {\bibinfo {volume} {304}},\ \bibinfo {pages} {L25} (\bibinfo {year}
  {1999})}\BibitemShut {NoStop}%
\bibitem [{\citenamefont {Annala}\ \emph {et~al.}(2020)\citenamefont {Annala},
  \citenamefont {Gorda}, \citenamefont {Kurkela}, \citenamefont
  {N{\"a}ttil{\"a}},\ and\ \citenamefont {Vuorinen}}]{annala2020evidence}%
  \BibitemOpen
  \bibfield  {author} {\bibinfo {author} {\bibfnamefont {E.}~\bibnamefont
  {Annala}}, \bibinfo {author} {\bibfnamefont {T.}~\bibnamefont {Gorda}},
  \bibinfo {author} {\bibfnamefont {A.}~\bibnamefont {Kurkela}}, \bibinfo
  {author} {\bibfnamefont {J.}~\bibnamefont {N{\"a}ttil{\"a}}}, \ and\ \bibinfo
  {author} {\bibfnamefont {A.}~\bibnamefont {Vuorinen}},\ }\href@noop {}
  {\bibfield  {journal} {\bibinfo  {journal} {Nature Physics}\ }\textbf
  {\bibinfo {volume} {16}},\ \bibinfo {pages} {907} (\bibinfo {year}
  {2020})}\BibitemShut {NoStop}%
\bibitem [{\citenamefont {McLerran}\ and\ \citenamefont
  {Pisarski}(2007)}]{McLerran:2007qj}%
  \BibitemOpen
  \bibfield  {author} {\bibinfo {author} {\bibfnamefont {L.}~\bibnamefont
  {McLerran}}\ and\ \bibinfo {author} {\bibfnamefont {R.~D.}\ \bibnamefont
  {Pisarski}},\ }\href {\doibase 10.1016/j.nuclphysa.2007.08.013} {\bibfield
  {journal} {\bibinfo  {journal} {Nucl. Phys. A}\ }\textbf {\bibinfo {volume}
  {796}},\ \bibinfo {pages} {83} (\bibinfo {year} {2007})},\ \Eprint
  {http://arxiv.org/abs/0706.2191} {arXiv:0706.2191 [hep-ph]} \BibitemShut
  {NoStop}%
\bibitem [{\citenamefont {McLerran}\ and\ \citenamefont
  {Reddy}(2019)}]{McLerran:2018hbz}%
  \BibitemOpen
  \bibfield  {author} {\bibinfo {author} {\bibfnamefont {L.}~\bibnamefont
  {McLerran}}\ and\ \bibinfo {author} {\bibfnamefont {S.}~\bibnamefont
  {Reddy}},\ }\href {\doibase 10.1103/PhysRevLett.122.122701} {\bibfield
  {journal} {\bibinfo  {journal} {Phys. Rev. Lett.}\ }\textbf {\bibinfo
  {volume} {122}},\ \bibinfo {pages} {122701} (\bibinfo {year} {2019})},\
  \Eprint {http://arxiv.org/abs/1811.12503} {arXiv:1811.12503 [nucl-th]}
  \BibitemShut {NoStop}%
\bibitem [{\citenamefont {Tolos}\ and\ \citenamefont
  {Fabbietti}(2020)}]{Tolos:2020aln}%
  \BibitemOpen
  \bibfield  {author} {\bibinfo {author} {\bibfnamefont {L.}~\bibnamefont
  {Tolos}}\ and\ \bibinfo {author} {\bibfnamefont {L.}~\bibnamefont
  {Fabbietti}},\ }\href {\doibase 10.1016/j.ppnp.2020.103770} {\bibfield
  {journal} {\bibinfo  {journal} {Prog. Part. Nucl. Phys.}\ }\textbf {\bibinfo
  {volume} {112}},\ \bibinfo {pages} {103770} (\bibinfo {year} {2020})},\
  \Eprint {http://arxiv.org/abs/2002.09223} {arXiv:2002.09223 [nucl-ex]}
  \BibitemShut {NoStop}%
\bibitem [{\citenamefont {Bombaci}(2017)}]{Bombaci:2016xzl}%
  \BibitemOpen
  \bibfield  {author} {\bibinfo {author} {\bibfnamefont {I.}~\bibnamefont
  {Bombaci}},\ }\href {\doibase 10.7566/JPSCP.17.101002} {\bibfield  {journal}
  {\bibinfo  {journal} {JPS Conf. Proc.}\ }\textbf {\bibinfo {volume} {17}},\
  \bibinfo {pages} {101002} (\bibinfo {year} {2017})},\ \Eprint
  {http://arxiv.org/abs/1601.05339} {arXiv:1601.05339 [nucl-th]} \BibitemShut
  {NoStop}%
\bibitem [{\citenamefont {Vida\~na}(2016)}]{Vidana:2015rsa}%
  \BibitemOpen
  \bibfield  {author} {\bibinfo {author} {\bibfnamefont {I.}~\bibnamefont
  {Vida\~na}},\ }\href {\doibase 10.1088/1742-6596/668/1/012031} {\bibfield
  {journal} {\bibinfo  {journal} {J. Phys. Conf. Ser.}\ }\textbf {\bibinfo
  {volume} {668}},\ \bibinfo {pages} {012031} (\bibinfo {year} {2016})},\
  \Eprint {http://arxiv.org/abs/1509.03587} {arXiv:1509.03587 [nucl-th]}
  \BibitemShut {NoStop}%
\bibitem [{\citenamefont {Skyrme}(1962)}]{skyrme1962unified}%
  \BibitemOpen
  \bibfield  {author} {\bibinfo {author} {\bibfnamefont {T.~H.~R.}\
  \bibnamefont {Skyrme}},\ }\href@noop {} {\bibfield  {journal} {\bibinfo
  {journal} {Nuclear Physics}\ }\textbf {\bibinfo {volume} {31}},\ \bibinfo
  {pages} {556} (\bibinfo {year} {1962})}\BibitemShut {NoStop}%
\bibitem [{\citenamefont {Adkins}\ \emph {et~al.}(1983)\citenamefont {Adkins},
  \citenamefont {Nappi},\ and\ \citenamefont {Witten}}]{adkins1983static}%
  \BibitemOpen
  \bibfield  {author} {\bibinfo {author} {\bibfnamefont {G.~S.}\ \bibnamefont
  {Adkins}}, \bibinfo {author} {\bibfnamefont {C.~R.}\ \bibnamefont {Nappi}}, \
  and\ \bibinfo {author} {\bibfnamefont {E.}~\bibnamefont {Witten}},\
  }\href@noop {} {\bibfield  {journal} {\bibinfo  {journal} {Nuclear Physics
  B}\ }\textbf {\bibinfo {volume} {228}},\ \bibinfo {pages} {552} (\bibinfo
  {year} {1983})}\BibitemShut {NoStop}%
\bibitem [{\citenamefont {Ding}\ and\ \citenamefont {Yan}(2007)}]{Ding:2007xi}%
  \BibitemOpen
  \bibfield  {author} {\bibinfo {author} {\bibfnamefont {G.-J.}\ \bibnamefont
  {Ding}}\ and\ \bibinfo {author} {\bibfnamefont {M.-L.}\ \bibnamefont {Yan}},\
  }\href {\doibase 10.1103/PhysRevC.75.034004} {\bibfield  {journal} {\bibinfo
  {journal} {Phys. Rev. C}\ }\textbf {\bibinfo {volume} {75}},\ \bibinfo
  {pages} {034004} (\bibinfo {year} {2007})},\ \Eprint
  {http://arxiv.org/abs/nucl-th/0702037} {arXiv:nucl-th/0702037} \BibitemShut
  {NoStop}%
\bibitem [{\citenamefont {Halcrow}\ and\ \citenamefont
  {Harland}(2020)}]{Halcrow:2020gbm}%
  \BibitemOpen
  \bibfield  {author} {\bibinfo {author} {\bibfnamefont {C.}~\bibnamefont
  {Halcrow}}\ and\ \bibinfo {author} {\bibfnamefont {D.}~\bibnamefont
  {Harland}},\ }\href {\doibase 10.1103/PhysRevLett.125.042501} {\bibfield
  {journal} {\bibinfo  {journal} {Phys. Rev. Lett.}\ }\textbf {\bibinfo
  {volume} {125}},\ \bibinfo {pages} {042501} (\bibinfo {year}
  {2020})}\BibitemShut {NoStop}%
\bibitem [{\citenamefont {Lau}\ and\ \citenamefont
  {Manton}(2014)}]{Lau:2014baa}%
  \BibitemOpen
  \bibfield  {author} {\bibinfo {author} {\bibfnamefont {P.~H.~C.}\
  \bibnamefont {Lau}}\ and\ \bibinfo {author} {\bibfnamefont {N.~S.}\
  \bibnamefont {Manton}},\ }\href {\doibase 10.1103/PhysRevLett.113.232503}
  {\bibfield  {journal} {\bibinfo  {journal} {Phys. Rev. Lett.}\ }\textbf
  {\bibinfo {volume} {113}},\ \bibinfo {pages} {232503} (\bibinfo {year}
  {2014})}\BibitemShut {NoStop}%
\bibitem [{\citenamefont {Halcrow}\ \emph {et~al.}(2017)\citenamefont
  {Halcrow}, \citenamefont {King},\ and\ \citenamefont
  {Manton}}]{Halcrow:2016spb}%
  \BibitemOpen
  \bibfield  {author} {\bibinfo {author} {\bibfnamefont {C.~J.}\ \bibnamefont
  {Halcrow}}, \bibinfo {author} {\bibfnamefont {C.}~\bibnamefont {King}}, \
  and\ \bibinfo {author} {\bibfnamefont {N.~S.}\ \bibnamefont {Manton}},\
  }\href {\doibase 10.1103/PhysRevC.95.031303} {\bibfield  {journal} {\bibinfo
  {journal} {Phys. Rev. C}\ }\textbf {\bibinfo {volume} {95}},\ \bibinfo
  {pages} {031303} (\bibinfo {year} {2017})}\BibitemShut {NoStop}%
\bibitem [{\citenamefont {Adam}\ \emph {et~al.}(2013)\citenamefont {Adam},
  \citenamefont {Naya}, \citenamefont {Sanchez-Guillen},\ and\ \citenamefont
  {Wereszczynski}}]{Adam:2013wya}%
  \BibitemOpen
  \bibfield  {author} {\bibinfo {author} {\bibfnamefont {C.}~\bibnamefont
  {Adam}}, \bibinfo {author} {\bibfnamefont {C.}~\bibnamefont {Naya}}, \bibinfo
  {author} {\bibfnamefont {J.}~\bibnamefont {Sanchez-Guillen}}, \ and\ \bibinfo
  {author} {\bibfnamefont {A.}~\bibnamefont {Wereszczynski}},\ }\href {\doibase
  10.1103/PhysRevLett.111.232501} {\bibfield  {journal} {\bibinfo  {journal}
  {Phys. Rev. Lett.}\ }\textbf {\bibinfo {volume} {111}},\ \bibinfo {pages}
  {232501} (\bibinfo {year} {2013})},\ \Eprint {http://arxiv.org/abs/1312.2960}
  {arXiv:1312.2960 [nucl-th]} \BibitemShut {NoStop}%
\bibitem [{\citenamefont {Gillard}\ \emph {et~al.}(2015)\citenamefont
  {Gillard}, \citenamefont {Harland},\ and\ \citenamefont
  {Speight}}]{Gillard:2015eia}%
  \BibitemOpen
  \bibfield  {author} {\bibinfo {author} {\bibfnamefont {M.}~\bibnamefont
  {Gillard}}, \bibinfo {author} {\bibfnamefont {D.}~\bibnamefont {Harland}}, \
  and\ \bibinfo {author} {\bibfnamefont {M.}~\bibnamefont {Speight}},\ }\href
  {\doibase 10.1016/j.nuclphysb.2015.04.005} {\bibfield  {journal} {\bibinfo
  {journal} {Nucl. Phys. B}\ }\textbf {\bibinfo {volume} {895}},\ \bibinfo
  {pages} {272} (\bibinfo {year} {2015})}\BibitemShut {NoStop}%
\bibitem [{\citenamefont {Gudnason}(2016)}]{Gudnason:2016mms}%
  \BibitemOpen
  \bibfield  {author} {\bibinfo {author} {\bibfnamefont {S.~B.}\ \bibnamefont
  {Gudnason}},\ }\href {\doibase 10.1103/PhysRevD.93.065048} {\bibfield
  {journal} {\bibinfo  {journal} {Phys. Rev. D}\ }\textbf {\bibinfo {volume}
  {93}},\ \bibinfo {pages} {065048} (\bibinfo {year} {2016})}\BibitemShut
  {NoStop}%
\bibitem [{\citenamefont {Naya}\ and\ \citenamefont
  {Sutcliffe}(2018)}]{Naya:2018kyi}%
  \BibitemOpen
  \bibfield  {author} {\bibinfo {author} {\bibfnamefont {C.}~\bibnamefont
  {Naya}}\ and\ \bibinfo {author} {\bibfnamefont {P.}~\bibnamefont
  {Sutcliffe}},\ }\href {\doibase 10.1103/PhysRevLett.121.232002} {\bibfield
  {journal} {\bibinfo  {journal} {Phys. Rev. Lett.}\ }\textbf {\bibinfo
  {volume} {121}},\ \bibinfo {pages} {232002} (\bibinfo {year}
  {2018})}\BibitemShut {NoStop}%
\bibitem [{\citenamefont {Adam}\ \emph {et~al.}(2010)\citenamefont {Adam},
  \citenamefont {Sánchez-Guillén},\ and\ \citenamefont
  {Wereszczyński}}]{Adam_2010}%
  \BibitemOpen
  \bibfield  {author} {\bibinfo {author} {\bibfnamefont {C.}~\bibnamefont
  {Adam}}, \bibinfo {author} {\bibfnamefont {J.}~\bibnamefont
  {Sánchez-Guillén}}, \ and\ \bibinfo {author} {\bibfnamefont
  {A.}~\bibnamefont {Wereszczyński}},\ }\href {\doibase
  10.1016/j.physletb.2010.06.025} {\bibfield  {journal} {\bibinfo  {journal}
  {Physics Letters B}\ }\textbf {\bibinfo {volume} {691}},\ \bibinfo {pages}
  {105–110} (\bibinfo {year} {2010})}\BibitemShut {NoStop}%
\bibitem [{\citenamefont {Naya}(2019)}]{naya2019neutron}%
  \BibitemOpen
  \bibfield  {author} {\bibinfo {author} {\bibfnamefont {C.}~\bibnamefont
  {Naya}},\ }\href@noop {} {\bibfield  {journal} {\bibinfo  {journal}
  {International Journal of Modern Physics E}\ }\textbf {\bibinfo {volume}
  {28}},\ \bibinfo {pages} {1930006} (\bibinfo {year} {2019})}\BibitemShut
  {NoStop}%
\bibitem [{\citenamefont {Adam}\ \emph {et~al.}(2020)\citenamefont {Adam},
  \citenamefont {Martín-Caro}, \citenamefont {Huidobro}, \citenamefont
  {V\'azquez},\ and\ \citenamefont {Wereszczynski}}]{Adam:2020yfv}%
  \BibitemOpen
  \bibfield  {author} {\bibinfo {author} {\bibfnamefont {C.}~\bibnamefont
  {Adam}}, \bibinfo {author} {\bibfnamefont {A.~G.}\ \bibnamefont
  {Martín-Caro}}, \bibinfo {author} {\bibfnamefont {M.}~\bibnamefont
  {Huidobro}}, \bibinfo {author} {\bibfnamefont {R.}~\bibnamefont {V\'azquez}},
  \ and\ \bibinfo {author} {\bibfnamefont {A.}~\bibnamefont {Wereszczynski}},\
  }\href {\doibase 10.1016/j.physletb.2020.135928} {\bibfield  {journal}
  {\bibinfo  {journal} {Phys. Lett. B}\ }\textbf {\bibinfo {volume} {811}},\
  \bibinfo {pages} {135928} (\bibinfo {year} {2020})}\BibitemShut {NoStop}%
\bibitem [{\citenamefont {Adam}\ \emph
  {et~al.}(2022{\natexlab{a}})\citenamefont {Adam}, \citenamefont
  {Martin-Caro}, \citenamefont {Huidobro}, \citenamefont {Vazquez},\ and\
  \citenamefont {Wereszczynski}}]{Adam:2021gbm}%
  \BibitemOpen
  \bibfield  {author} {\bibinfo {author} {\bibfnamefont {C.}~\bibnamefont
  {Adam}}, \bibinfo {author} {\bibfnamefont {A.~G.}\ \bibnamefont
  {Martin-Caro}}, \bibinfo {author} {\bibfnamefont {M.}~\bibnamefont
  {Huidobro}}, \bibinfo {author} {\bibfnamefont {R.}~\bibnamefont {Vazquez}}, \
  and\ \bibinfo {author} {\bibfnamefont {A.}~\bibnamefont {Wereszczynski}},\
  }\href {\doibase 10.1103/PhysRevD.105.074019} {\bibfield  {journal} {\bibinfo
   {journal} {Phys. Rev. D}\ }\textbf {\bibinfo {volume} {105}},\ \bibinfo
  {pages} {074019} (\bibinfo {year} {2022}{\natexlab{a}})},\ \Eprint
  {http://arxiv.org/abs/2109.13946} {arXiv:2109.13946 [hep-th]} \BibitemShut
  {NoStop}%
\bibitem [{\citenamefont {Adam}\ \emph
  {et~al.}(2022{\natexlab{b}})\citenamefont {Adam}, \citenamefont
  {Mart\'\i{}n-Caro}, \citenamefont {Huidobro}, \citenamefont {V\'azquez},\
  and\ \citenamefont {Wereszczynski}}]{Adam:2022aes}%
  \BibitemOpen
  \bibfield  {author} {\bibinfo {author} {\bibfnamefont {C.}~\bibnamefont
  {Adam}}, \bibinfo {author} {\bibfnamefont {A.~G.}\ \bibnamefont
  {Mart\'\i{}n-Caro}}, \bibinfo {author} {\bibfnamefont {M.}~\bibnamefont
  {Huidobro}}, \bibinfo {author} {\bibfnamefont {R.}~\bibnamefont {V\'azquez}},
  \ and\ \bibinfo {author} {\bibfnamefont {A.}~\bibnamefont {Wereszczynski}},\
  }\href@noop {} {\  (\bibinfo {year} {2022}{\natexlab{b}})},\ \Eprint
  {http://arxiv.org/abs/2202.00953} {arXiv:2202.00953 [nucl-th]} \BibitemShut
  {NoStop}%
\bibitem [{\citenamefont {Adam}\ \emph {et~al.}(2015)\citenamefont {Adam},
  \citenamefont {Naya}, \citenamefont {Sanchez-Guillen}, \citenamefont
  {Vazquez},\ and\ \citenamefont {Wereszczynski}}]{Adam_2015a}%
  \BibitemOpen
  \bibfield  {author} {\bibinfo {author} {\bibfnamefont {C.}~\bibnamefont
  {Adam}}, \bibinfo {author} {\bibfnamefont {C.}~\bibnamefont {Naya}}, \bibinfo
  {author} {\bibfnamefont {J.}~\bibnamefont {Sanchez-Guillen}}, \bibinfo
  {author} {\bibfnamefont {R.}~\bibnamefont {Vazquez}}, \ and\ \bibinfo
  {author} {\bibfnamefont {A.}~\bibnamefont {Wereszczynski}},\ }\href {\doibase
  10.1016/j.physletb.2015.01.027} {\bibfield  {journal} {\bibinfo  {journal}
  {Physics Letters B}\ }\textbf {\bibinfo {volume} {742}},\ \bibinfo {pages}
  {136–142} (\bibinfo {year} {2015})}\BibitemShut {NoStop}%
\bibitem [{\citenamefont {Callan}\ and\ \citenamefont
  {Klebanov}(1985)}]{callan1985bound}%
  \BibitemOpen
  \bibfield  {author} {\bibinfo {author} {\bibfnamefont {C.~G.}\ \bibnamefont
  {Callan}}\ and\ \bibinfo {author} {\bibfnamefont {I.}~\bibnamefont
  {Klebanov}},\ }\href@noop {} {\bibfield  {journal} {\bibinfo  {journal}
  {Nuclear Physics B}\ }\textbf {\bibinfo {volume} {262}},\ \bibinfo {pages}
  {365} (\bibinfo {year} {1985})}\BibitemShut {NoStop}%
\bibitem [{\citenamefont {Klebanov}(1990)}]{Klebanov1990}%
  \BibitemOpen
  \bibfield  {author} {\bibinfo {author} {\bibfnamefont {I.}~\bibnamefont
  {Klebanov}},\ }\enquote {\bibinfo {title} {Strangeness in the skyrme
  model},}\ in\ \href {\doibase 10.1007/978-1-4684-1336-6_8} {\emph {\bibinfo
  {booktitle} {Hadrons and Hadronic Matter}}},\ \bibinfo {editor} {edited by\
  \bibinfo {editor} {\bibfnamefont {D.}~\bibnamefont {Vautherin}}, \bibinfo
  {editor} {\bibfnamefont {F.}~\bibnamefont {Lenz}}, \ and\ \bibinfo {editor}
  {\bibfnamefont {J.~W.}\ \bibnamefont {Negele}}}\ (\bibinfo  {publisher}
  {Springer US},\ \bibinfo {address} {Boston, MA},\ \bibinfo {year} {1990})\
  pp.\ \bibinfo {pages} {223--262}\BibitemShut {NoStop}%
\bibitem [{\citenamefont {Fiorella~Burgio}\ and\ \citenamefont
  {Fantina}(2018)}]{FiorellaFantina:2018dga}%
  \BibitemOpen
  \bibfield  {author} {\bibinfo {author} {\bibfnamefont {G.}~\bibnamefont
  {Fiorella~Burgio}}\ and\ \bibinfo {author} {\bibfnamefont {A.~F.}\
  \bibnamefont {Fantina}},\ }\href {\doibase 10.1007/978-3-319-97616-7_6}
  {\bibfield  {journal} {\bibinfo  {journal} {Astrophys. Space Sci. Libr.}\
  }\textbf {\bibinfo {volume} {457}},\ \bibinfo {pages} {255} (\bibinfo {year}
  {2018})}\BibitemShut {NoStop}%
\bibitem [{\citenamefont {Essick}\ \emph {et~al.}(2021)\citenamefont {Essick},
  \citenamefont {Landry}, \citenamefont {Schwenk},\ and\ \citenamefont
  {Tews}}]{Landry:2021ezp}%
  \BibitemOpen
  \bibfield  {author} {\bibinfo {author} {\bibfnamefont {R.}~\bibnamefont
  {Essick}}, \bibinfo {author} {\bibfnamefont {P.}~\bibnamefont {Landry}},
  \bibinfo {author} {\bibfnamefont {A.}~\bibnamefont {Schwenk}}, \ and\
  \bibinfo {author} {\bibfnamefont {I.}~\bibnamefont {Tews}},\ }\href {\doibase
  10.1103/PhysRevC.104.065804} {\bibfield  {journal} {\bibinfo  {journal}
  {Phys. Rev. C}\ }\textbf {\bibinfo {volume} {104}},\ \bibinfo {pages}
  {065804} (\bibinfo {year} {2021})}\BibitemShut {NoStop}%
\bibitem [{\citenamefont {Tang}\ \emph {et~al.}(2021)\citenamefont {Tang},
  \citenamefont {Jiang}, \citenamefont {Han}, \citenamefont {Fan},\ and\
  \citenamefont {Wei}}]{Tang:2021snt}%
  \BibitemOpen
  \bibfield  {author} {\bibinfo {author} {\bibfnamefont {S.-P.}\ \bibnamefont
  {Tang}}, \bibinfo {author} {\bibfnamefont {J.-L.}\ \bibnamefont {Jiang}},
  \bibinfo {author} {\bibfnamefont {M.-Z.}\ \bibnamefont {Han}}, \bibinfo
  {author} {\bibfnamefont {Y.-Z.}\ \bibnamefont {Fan}}, \ and\ \bibinfo
  {author} {\bibfnamefont {D.-M.}\ \bibnamefont {Wei}},\ }\href {\doibase
  10.1103/PhysRevD.104.063032} {\bibfield  {journal} {\bibinfo  {journal}
  {Phys. Rev. D}\ }\textbf {\bibinfo {volume} {104}},\ \bibinfo {pages}
  {063032} (\bibinfo {year} {2021})}\BibitemShut {NoStop}%
\bibitem [{\citenamefont {de~Tovar}\ \emph {et~al.}(2021)\citenamefont
  {de~Tovar}, \citenamefont {Ferreira},\ and\ \citenamefont
  {Provid\^encia}}]{deTovar:2021sjo}%
  \BibitemOpen
  \bibfield  {author} {\bibinfo {author} {\bibfnamefont {P.~B.}\ \bibnamefont
  {de~Tovar}}, \bibinfo {author} {\bibfnamefont {M.}~\bibnamefont {Ferreira}},
  \ and\ \bibinfo {author} {\bibfnamefont {C.}~\bibnamefont {Provid\^encia}},\
  }\href {\doibase 10.1103/PhysRevD.104.123036} {\bibfield  {journal} {\bibinfo
   {journal} {Phys. Rev. D}\ }\textbf {\bibinfo {volume} {104}},\ \bibinfo
  {pages} {123036} (\bibinfo {year} {2021})}\BibitemShut {NoStop}%
\bibitem [{\citenamefont {Gil}\ \emph {et~al.}(2021)\citenamefont {Gil},
  \citenamefont {Papakonstantinou},\ and\ \citenamefont {Hyun}}]{Gil:2021ols}%
  \BibitemOpen
  \bibfield  {author} {\bibinfo {author} {\bibfnamefont {H.}~\bibnamefont
  {Gil}}, \bibinfo {author} {\bibfnamefont {P.}~\bibnamefont
  {Papakonstantinou}}, \ and\ \bibinfo {author} {\bibfnamefont {C.~H.}\
  \bibnamefont {Hyun}},\ }\href@noop {} {\  (\bibinfo {year} {2021})},\ \Eprint
  {http://arxiv.org/abs/2110.09802} {arXiv:2110.09802 [nucl-th]} \BibitemShut
  {NoStop}%
\bibitem [{\citenamefont {Li}\ \emph {et~al.}(2021)\citenamefont {Li},
  \citenamefont {Cai}, \citenamefont {Xie},\ and\ \citenamefont
  {Zhang}}]{Li:2021thg}%
  \BibitemOpen
  \bibfield  {author} {\bibinfo {author} {\bibfnamefont {B.-A.}\ \bibnamefont
  {Li}}, \bibinfo {author} {\bibfnamefont {B.-J.}\ \bibnamefont {Cai}},
  \bibinfo {author} {\bibfnamefont {W.-J.}\ \bibnamefont {Xie}}, \ and\
  \bibinfo {author} {\bibfnamefont {N.-B.}\ \bibnamefont {Zhang}},\ }\href
  {\doibase 10.3390/universe7060182} {\bibfield  {journal} {\bibinfo  {journal}
  {Universe}\ }\textbf {\bibinfo {volume} {7}},\ \bibinfo {pages} {182}
  (\bibinfo {year} {2021})}\BibitemShut {NoStop}%
\bibitem [{\citenamefont {Klebanov}(1985)}]{klebanov1985nuclear}%
  \BibitemOpen
  \bibfield  {author} {\bibinfo {author} {\bibfnamefont {I.}~\bibnamefont
  {Klebanov}},\ }\href@noop {} {\bibfield  {journal} {\bibinfo  {journal}
  {Nuclear Physics B}\ }\textbf {\bibinfo {volume} {262}},\ \bibinfo {pages}
  {133} (\bibinfo {year} {1985})}\BibitemShut {NoStop}%
\bibitem [{\citenamefont {Blom}\ \emph {et~al.}(1989)\citenamefont {Blom},
  \citenamefont {Dannbom},\ and\ \citenamefont {Riska}}]{Blom1989HyperonsAB}%
  \BibitemOpen
  \bibfield  {author} {\bibinfo {author} {\bibfnamefont {U.}~\bibnamefont
  {Blom}}, \bibinfo {author} {\bibfnamefont {K.}~\bibnamefont {Dannbom}}, \
  and\ \bibinfo {author} {\bibfnamefont {D.}~\bibnamefont {Riska}},\
  }\href@noop {} {\bibfield  {journal} {\bibinfo  {journal} {Nuclear Physics}\
  }\textbf {\bibinfo {volume} {493}},\ \bibinfo {pages} {384} (\bibinfo {year}
  {1989})}\BibitemShut {NoStop}%
\bibitem [{\citenamefont {Nyman}\ and\ \citenamefont
  {Riska}(1990)}]{nyman1990low}%
  \BibitemOpen
  \bibfield  {author} {\bibinfo {author} {\bibfnamefont {E.~M.}\ \bibnamefont
  {Nyman}}\ and\ \bibinfo {author} {\bibfnamefont {D.}~\bibnamefont {Riska}},\
  }\href@noop {} {\bibfield  {journal} {\bibinfo  {journal} {Reports on
  Progress in Physics}\ }\textbf {\bibinfo {volume} {53}},\ \bibinfo {pages}
  {1137} (\bibinfo {year} {1990})}\BibitemShut {NoStop}%
\bibitem [{\citenamefont {Schmitt}(2010)}]{Schmitt:2010pn}%
  \BibitemOpen
  \bibfield  {author} {\bibinfo {author} {\bibfnamefont {A.}~\bibnamefont
  {Schmitt}},\ }\href {\doibase 10.1007/978-3-642-12866-0} {\emph {\bibinfo
  {title} {{Dense matter in compact stars: A pedagogical introduction}}}},\
  Vol.\ \bibinfo {volume} {811}\ (\bibinfo {year} {2010})\ \Eprint
  {http://arxiv.org/abs/1001.3294} {arXiv:1001.3294 [astro-ph.SR]} \BibitemShut
  {NoStop}%
\bibitem [{Note1()}]{Note1}%
  \BibitemOpen
  \bibinfo {note} {Actually, that the charged (in particular, the negatively
  charged) kaons will condense first is true in our approach (whenever $\mu
  _e>0$), since the chemical potential associated to neutral kaons is zero, so
  that the onset of neutral kaon condensation is given by
  $m^*_K=0$.}\BibitemShut {Stop}%
\bibitem [{Note2()}]{Note2}%
  \BibitemOpen
  \bibinfo {note} {Remember that we are using the mostly minus convention for
  the metric signature.}\BibitemShut {Stop}%
\bibitem [{\citenamefont {Park}\ \emph {et~al.}(2019)\citenamefont {Park},
  \citenamefont {Paeng},\ and\ \citenamefont {Vento}}]{Park:2019bmi}%
  \BibitemOpen
  \bibfield  {author} {\bibinfo {author} {\bibfnamefont {B.-Y.}\ \bibnamefont
  {Park}}, \bibinfo {author} {\bibfnamefont {W.-G.}\ \bibnamefont {Paeng}}, \
  and\ \bibinfo {author} {\bibfnamefont {V.}~\bibnamefont {Vento}},\ }\href
  {\doibase 10.1016/j.nuclphysa.2019.06.010} {\bibfield  {journal} {\bibinfo
  {journal} {Nucl. Phys. A}\ }\textbf {\bibinfo {volume} {989}},\ \bibinfo
  {pages} {231} (\bibinfo {year} {2019})},\ \Eprint
  {http://arxiv.org/abs/1904.04483} {arXiv:1904.04483 [hep-ph]} \BibitemShut
  {NoStop}%
\bibitem [{\citenamefont {Canfora}\ \emph {et~al.}(2020)\citenamefont
  {Canfora}, \citenamefont {Lagos},\ and\ \citenamefont
  {Vera}}]{Canfora:2020kyj}%
  \BibitemOpen
  \bibfield  {author} {\bibinfo {author} {\bibfnamefont {F.}~\bibnamefont
  {Canfora}}, \bibinfo {author} {\bibfnamefont {M.}~\bibnamefont {Lagos}}, \
  and\ \bibinfo {author} {\bibfnamefont {A.}~\bibnamefont {Vera}},\ }\href
  {\doibase 10.1140/epjc/s10052-020-8275-1} {\bibfield  {journal} {\bibinfo
  {journal} {Eur. Phys. J. C}\ }\textbf {\bibinfo {volume} {80}},\ \bibinfo
  {pages} {697} (\bibinfo {year} {2020})},\ \Eprint
  {http://arxiv.org/abs/2007.11543} {arXiv:2007.11543 [hep-th]} \BibitemShut
  {NoStop}%
\bibitem [{\citenamefont {Glendenning}(1992)}]{Glendenning:1992vb}%
  \BibitemOpen
  \bibfield  {author} {\bibinfo {author} {\bibfnamefont {N.~K.}\ \bibnamefont
  {Glendenning}},\ }\href {\doibase 10.1103/PhysRevD.46.1274} {\bibfield
  {journal} {\bibinfo  {journal} {Phys. Rev. D}\ }\textbf {\bibinfo {volume}
  {46}},\ \bibinfo {pages} {1274} (\bibinfo {year} {1992})}\BibitemShut
  {NoStop}%
\bibitem [{\citenamefont {Bhattacharyya}\ \emph {et~al.}(2010)\citenamefont
  {Bhattacharyya}, \citenamefont {Mishustin},\ and\ \citenamefont
  {Greiner}}]{Bhattacharyya:2009fg}%
  \BibitemOpen
  \bibfield  {author} {\bibinfo {author} {\bibfnamefont {A.}~\bibnamefont
  {Bhattacharyya}}, \bibinfo {author} {\bibfnamefont {I.~N.}\ \bibnamefont
  {Mishustin}}, \ and\ \bibinfo {author} {\bibfnamefont {W.}~\bibnamefont
  {Greiner}},\ }\href {\doibase 10.1088/0954-3899/37/2/025201} {\bibfield
  {journal} {\bibinfo  {journal} {J. Phys. G}\ }\textbf {\bibinfo {volume}
  {37}},\ \bibinfo {pages} {025201} (\bibinfo {year} {2010})},\ \Eprint
  {http://arxiv.org/abs/0905.0352} {arXiv:0905.0352 [nucl-th]} \BibitemShut
  {NoStop}%
\bibitem [{\citenamefont {Sharma}\ \emph {et~al.}(2015)\citenamefont {Sharma},
  \citenamefont {Centelles}, \citenamefont {Vi\~nas}, \citenamefont {Baldo},\
  and\ \citenamefont {Burgio}}]{Sharma:2015bna}%
  \BibitemOpen
  \bibfield  {author} {\bibinfo {author} {\bibfnamefont {B.~K.}\ \bibnamefont
  {Sharma}}, \bibinfo {author} {\bibfnamefont {M.}~\bibnamefont {Centelles}},
  \bibinfo {author} {\bibfnamefont {X.}~\bibnamefont {Vi\~nas}}, \bibinfo
  {author} {\bibfnamefont {M.}~\bibnamefont {Baldo}}, \ and\ \bibinfo {author}
  {\bibfnamefont {G.~F.}\ \bibnamefont {Burgio}},\ }\href {\doibase
  10.1051/0004-6361/201526642} {\bibfield  {journal} {\bibinfo  {journal}
  {Astron. Astrophys.}\ }\textbf {\bibinfo {volume} {584}},\ \bibinfo {pages}
  {A103} (\bibinfo {year} {2015})}\BibitemShut {NoStop}%
\bibitem [{\citenamefont {Abbott}\ \emph {et~al.}(2017)\citenamefont {Abbott}
  \emph {et~al.}}]{Abbott_2017}%
  \BibitemOpen
  \bibfield  {author} {\bibinfo {author} {\bibfnamefont {B.~P.}\ \bibnamefont
  {Abbott}} \emph {et~al.} (\bibinfo {collaboration} {LIGO , VIRGO}),\ }\href
  {\doibase 10.1103/PhysRevLett.119.161101} {\bibfield  {journal} {\bibinfo
  {journal} {Phys. Rev. Lett.}\ }\textbf {\bibinfo {volume} {119}},\ \bibinfo
  {pages} {161101} (\bibinfo {year} {2017})}\BibitemShut {NoStop}%
\bibitem [{\citenamefont {Abbott}\ \emph {et~al.}(2020)\citenamefont {Abbott}
  \emph {et~al.}}]{Abbott:2020uma}%
  \BibitemOpen
  \bibfield  {author} {\bibinfo {author} {\bibfnamefont {B.}~\bibnamefont
  {Abbott}} \emph {et~al.} (\bibinfo {collaboration} {LIGO Scientific,
  Virgo}),\ }\href {\doibase 10.3847/2041-8213/ab75f5} {\bibfield  {journal}
  {\bibinfo  {journal} {Astrophys. J. Lett.}\ }\textbf {\bibinfo {volume}
  {892}},\ \bibinfo {pages} {L3} (\bibinfo {year} {2020})}\BibitemShut
  {NoStop}%
\bibitem [{\citenamefont {Miller}\ \emph {et~al.}(2019)\citenamefont {Miller}
  \emph {et~al.}}]{Miller:2019cac}%
  \BibitemOpen
  \bibfield  {author} {\bibinfo {author} {\bibfnamefont {M.~C.}\ \bibnamefont
  {Miller}} \emph {et~al.},\ }\href {\doibase 10.3847/2041-8213/ab50c5}
  {\bibfield  {journal} {\bibinfo  {journal} {Astrophys. J. Lett.}\ }\textbf
  {\bibinfo {volume} {887}},\ \bibinfo {pages} {L24} (\bibinfo {year}
  {2019})},\ \Eprint {http://arxiv.org/abs/1912.05705} {arXiv:1912.05705
  [astro-ph.HE]} \BibitemShut {NoStop}%
\bibitem [{\citenamefont {Riley}\ \emph {et~al.}(2021)\citenamefont {Riley}
  \emph {et~al.}}]{Riley:2021pdl}%
  \BibitemOpen
  \bibfield  {author} {\bibinfo {author} {\bibfnamefont {T.~E.}\ \bibnamefont
  {Riley}} \emph {et~al.},\ }\href {\doibase 10.3847/2041-8213/ac0a81}
  {\bibfield  {journal} {\bibinfo  {journal} {Astrophys. J. Lett.}\ }\textbf
  {\bibinfo {volume} {918}},\ \bibinfo {pages} {L27} (\bibinfo {year}
  {2021})}\BibitemShut {NoStop}%
\bibitem [{\citenamefont {Altiparmak}\ \emph {et~al.}(2022)\citenamefont
  {Altiparmak}, \citenamefont {Ecker},\ and\ \citenamefont
  {Rezzolla}}]{Altiparmak:2022bke}%
  \BibitemOpen
  \bibfield  {author} {\bibinfo {author} {\bibfnamefont {S.}~\bibnamefont
  {Altiparmak}}, \bibinfo {author} {\bibfnamefont {C.}~\bibnamefont {Ecker}}, \
  and\ \bibinfo {author} {\bibfnamefont {L.}~\bibnamefont {Rezzolla}},\
  }\href@noop {} {\  (\bibinfo {year} {2022})},\ \Eprint
  {http://arxiv.org/abs/2203.14974} {arXiv:2203.14974 [astro-ph.HE]}
  \BibitemShut {NoStop}%
\bibitem [{\citenamefont {Westerberg}(1995)}]{Westerberg:1994hu}%
  \BibitemOpen
  \bibfield  {author} {\bibinfo {author} {\bibfnamefont {K.~M.}\ \bibnamefont
  {Westerberg}},\ }\href {\doibase 10.1103/PhysRevD.51.5030} {\bibfield
  {journal} {\bibinfo  {journal} {Phys. Rev. D}\ }\textbf {\bibinfo {volume}
  {51}},\ \bibinfo {pages} {5030} (\bibinfo {year} {1995})},\ \Eprint
  {http://arxiv.org/abs/hep-ph/9411430} {arXiv:hep-ph/9411430} \BibitemShut
  {NoStop}%
\bibitem [{\citenamefont {Park}\ \emph {et~al.}(2010)\citenamefont {Park},
  \citenamefont {Kim},\ and\ \citenamefont {Rho}}]{Park:2009mg}%
  \BibitemOpen
  \bibfield  {author} {\bibinfo {author} {\bibfnamefont {B.-Y.}\ \bibnamefont
  {Park}}, \bibinfo {author} {\bibfnamefont {J.-I.}\ \bibnamefont {Kim}}, \
  and\ \bibinfo {author} {\bibfnamefont {M.}~\bibnamefont {Rho}},\ }\href
  {\doibase 10.1103/PhysRevC.81.035203} {\bibfield  {journal} {\bibinfo
  {journal} {Phys. Rev. C}\ }\textbf {\bibinfo {volume} {81}},\ \bibinfo
  {pages} {035203} (\bibinfo {year} {2010})},\ \Eprint
  {http://arxiv.org/abs/0912.3213} {arXiv:0912.3213 [hep-ph]} \BibitemShut
  {NoStop}%
\bibitem [{\citenamefont {Hong}\ \emph {et~al.}(2019)\citenamefont {Hong},
  \citenamefont {Yakhshiev},\ and\ \citenamefont {Kim}}]{Hong:2018sqa}%
  \BibitemOpen
  \bibfield  {author} {\bibinfo {author} {\bibfnamefont {K.-H.}\ \bibnamefont
  {Hong}}, \bibinfo {author} {\bibfnamefont {U.}~\bibnamefont {Yakhshiev}}, \
  and\ \bibinfo {author} {\bibfnamefont {H.-C.}\ \bibnamefont {Kim}},\ }\href
  {\doibase 10.1103/PhysRevC.99.035212} {\bibfield  {journal} {\bibinfo
  {journal} {Phys. Rev. C}\ }\textbf {\bibinfo {volume} {99}},\ \bibinfo
  {pages} {035212} (\bibinfo {year} {2019})},\ \Eprint
  {http://arxiv.org/abs/1806.06504} {arXiv:1806.06504 [nucl-th]} \BibitemShut
  {NoStop}%
\bibitem [{Note3()}]{Note3}%
  \BibitemOpen
  \bibinfo {note} {The result is of course independent of such extension,
  because $\pi _4(SU(3))$ vanishes.}\BibitemShut {Stop}%
\end{thebibliography}

%

\end{document}